\documentclass[prx,twocolumn,preprintnumbers,amsmath,amssymb,superscriptaddress]{revtex4}

\usepackage{svg}
\usepackage{color}
\usepackage{transparent}
\usepackage{soul}
\usepackage{graphicx}
\usepackage{dcolumn}
\usepackage{bm}
\usepackage{hyperref}
\usepackage{calc}
\usepackage{physics}

\begin{document}

\title{Majorana braiding gates for topological superconductors in a one dimensional geometry}

\author{Marek Narozniak}
\affiliation{New York University Shanghai, 1555 Century Ave, Pudong, Shanghai 200122, China}
\affiliation{Department of Physics, New York University, New York, NY, 10003, USA.}

\author{Matthieu Dartiailh}
\affiliation{Department of Physics, New York University, New York, NY, 10003, USA.}

\author{Jonathan P. Dowling}
\affiliation{Hearne Institute for Theoretical Physics, Department of Physics and Astronomy,
Louisiana State University, Baton Rouge, Louisiana 70803, USA}

\author{Javad Shabani}
\affiliation{Department of Physics, New York University, New York, NY, 10003, USA.}

\author{Tim Byrnes}
\email{tim.byrnes@nyu.edu}
\affiliation{New York University Shanghai, 1555 Century Ave, Pudong, Shanghai 200122, China}
\affiliation{State Key Laboratory of Precision Spectroscopy, School of Physical and Material Sciences, East China Normal University, Shanghai 200062, China}
\affiliation{NYU-ECNU Institute of Physics at NYU Shanghai, 3663 Zhongshan Road North, Shanghai 200062, China}
\affiliation{National Institute of Informatics, 2-1-2 Hitotsubashi, Chiyoda-ku, Tokyo 101-8430, Japan}
\affiliation{Department of Physics, New York University, New York, NY, 10003, USA.}

\date{\today}

\begin{abstract}
We propose and analyze a physical system capable of performing topological quantum computation with Majorana zero modes (MZM) in a one-dimensional topological superconductor (1DTS). One of the leading methods to realize quantum gates in 1DTS is to use T-junctions, which allows one to maneuver MZMs such as to achieve braiding.  In this paper, we propose a scheme that is in a purely one-dimensional geometry and does not require T-junctions, instead replacing it with an auxiliary qubit. We show that this allows one to perform one and two logical qubit $ Z $ rotations.  We first design a topologically protected logical $Z$-gate based entirely on local interactions within the 1DTS. Using an auxiliary qubit coupled to the topological superconductors, we extend the $Z$-gate to single and multiqubit arbitrary rotations with partial topological protection. Finally, to perform universal quantum computing, we introduce a scheme for performing arbitrary unitary rotations, albeit without topological protection.  We develop a formalism based on unitary braids which creates transitions between different topological phases of the 1DTS system.
The unitary formalism can be simply converted to an equivalent adiabatic scheme, which we numerically simulate and show that high fidelity operations should be possible with reasonable parameters.  
\end{abstract}

\maketitle

\section{Introduction}

Topological states of matter are an attractive medium for achieving a fault-tolerant model of quantum computation, based on the topological properties of anyons  \cite{kitaev2003fault,freedman2003topological,preskill2004topological,nayak2008non,pachos2012introduction,sarma2015majorana,lahtinen2017short}. This paradigm of quantum computing is a natural form of implementing quantum error correction \cite{Lahtinen2017ASI, freedman2003topological}, making the system --- as it scales up --- resistant to small perturbations and errors. The interchange of anyons is commonly referred as braiding \cite{kitaev2001unpaired,alicea2012new,leijnse2012introduction,beenakker2013search,sarma2015majorana,lutchyn2018majorana} and it remains immune to errors as long as the topology of the braiding path is not changed. If the particles that are being interchanged are non-Abelian anyons, such as Majorana Zero Modes (MZM), then their interchange can be used for performing quantum computation. Numerous proposals for topological quantum computing using Abelian anyons also exist, based on methods such as introducing dislocations in lattices \cite{You2013}. One of the sources of error protection is the energy gap between the subspaces for the logical states and the error states. In this sense, the topological description of quantum information becomes a particular way of storing and manipulating quantum information in a fault-tolerant way \cite{shor1996fault,gottesman1998theory,steane1999efficient,aharonov1999fault,preskill1998fault,gottesman2010introduction,devitt2013quantum,moxley2016sagnac,campbell2017roads}.  


Numerous platforms for observing and manipulating anyons with non-Abelian statistics have been proposed. Many possible implementations are based upon Fractional Quantum Hall Effect (FQHE) systems which have already been experimentally observed \cite{willett1987observation, Goldman1995, Saminadayar1997, willett2013magnetic, willett2009measurement}, although the direct observation of anyons remain elusive. Moreover, FQHE in alternative materials such as quantum magnets potentially opens the possibility of topological quantum computing at relatively high temperatures \cite{Kasahara2018,byrnes2015quantum,chen2019skyrmion}. Experimental proposals for qubits based on FQHE have also been suggested \cite{sarma2005topologically} as well as theoretical studies for achieving FQHE without superconductivity \cite{Wu2017}. Another promising candidate for a physical system that could implement topological quantum computation are one-dimensional topological superconductors (1DTS) such as 
nanowire semiconductor-superconductor heterostructures \cite{lutchyn2010majorana, oreg2010helical, mourik2012signatures, vanHeck2016, Zhang2018} or carbon nanotubes \cite{SanJose2015, Marganska2018, Lesser_2020}. Recently experimental evidence for zero-energy delocalized states on the wire ends was reported \cite{Vaitieknas2020}. A detailed state of the art on semiconductor-superconductor heterostructures and is provided in Ref. \cite{Lutchyn2018, Aguado2020, mayer2019phase}.


One of the most important considerations in designing a quantum computer is a robust way of performing quantum gates. In the case of topological quantum computing, this means designing a method of performing braiding of one or more anyons storing the quantum information. One of the best-known methods that has been proposed based on the T-junction geometry \cite{Alicea2011TJunction} consisting of assembling a system of three 1DTS with controllable coupling forming a characteristic shape of letter ``T''. Such a system has been shown to be able to swap any two MZMs by a suitable sequence of operations, and is able to maintain their delocalized nature. A disadvantage of this scheme is the difficulty of growing a heterostructure in this shape. A method to eliminate the T-junction, by replacing it with a auxiliary qubit was proposed in Ref. \cite{Backens2017}.  The auxiliary qubit, referred to as the coupler, is capable of performing an arbitrary single qubit $Z$-rotation in the logical space. The coupling-induced braid was only provided for a single strip of  1DTS, interacting the MZMs on the edges. This corresponds to a logical single qubit gate encoded by the MZMs, and no two logical qubit braiding was provided. Additionally, the single qubit gate only performed a rotation about the $Z$-axis, and no other types of gates were given.



\begin{figure}[t]
\includegraphics[width=\linewidth]{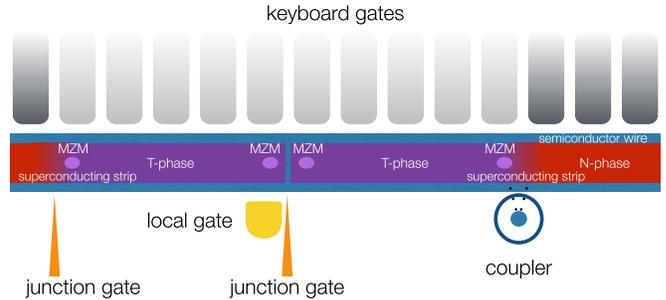}
\caption{An example of a controllable one-dimensional topological superconductor (1DTS) system that is considered in this work. A superconducting strip is placed on a semiconductor wire, which have regions in the topological phase ($T$-phase) or a normal phase ($N$-phase). Keyboard gates locally adjust the chemical potential, which changes the phase of the fermions in the superconducting strip. Junction gates are used to break the strips into regions, which at the endpoints have Majorana Zero Modes (MZMs), encoding the logical states.  A coupler, which consists of a qubit coupled to two local sites allows for braiding operations to be performed (see Sec. \ref{section_arbitrary_z_rotation_multi}).  Operations to control the local site are present to perform universal qubit gates (see Sec. \ref{section_arbitrary_qubit_rotation}). See Ref. \cite{Zhang2018, Aguado2020} for an experimentally implemented scheme with a similar configuration.}\label{processor_design}
\end{figure}

In this work, we propose and analyze protocols for achieving braiding gates for MZMs in a purely one-dimensional topology, extending the protocol of Ref. \cite{Backens2017} to multiple logical qubits. In Fig. \ref{processor_design}, we provide an overall sketch of the  one-dimensional quantum system considered in this paper. In our approach, the logical qubits are represented by separated regions of the 1DTS in the topological phases. Braiding operations are achieved by moving the topological regimes in the chain, with the aid of an auxiliary qubit (the coupler), which allows for control of the logical states.  Control gates such as the keyboard and junction gates allow for moving and manipulating the topological regions within the 1DTS, which results in logical operations. 
We introduce protocols allowing to perform a topologically protected logical $Z$-gate and partially protected arbitrary unitary rotations around the $Z$-axis for any number of topological qubits. We also provide an another non-topological braiding scheme for logical space rotation around a different axis making the braiding protocol capable of performing a universal quantum computation. 
One of the features of our work is that we introduce a unitary formalism describing the phase transitions between topologically trivial and topological regime. Understanding the unitary description of phase transition can potentially help design new logical gates and that is also how we used it in this work. 

\section{Majoranas in one-dimensional topological superconductors}

Before introducing our protocols for braiding MZMs in 1DTS, we describe the basics of the physical system. We start by reviewing fermion and Majorana operators and their properties, and introducing the Kitaev chain Hamiltonian modeling the 1DTS system. We will distinguish its phases and transitions between them and describe them using the unitary conversion operator. Finally, we will explain how unitary braiding is performed and explain how the logical space is defined in terms of the topological states that represent the quantum information.

\subsection{Majorana and quasifermion operator}

We start by describing the operators that describe the particles that comprise the system. Our system is a chain of mobile fermions that interact with a Bardeen-Cooper-Schrieffer (BCS) pairing interaction. The bare underlying fermions have fermionic creation and annihilation operators $a_n^\dagger$ and $a_n$ that satisfy the following properties
\begin{align}
 \{a_n, a_m\} &= 0 \label{fermions_anticommute_prop_1}\\
 \{a_n, a_m^\dagger\} &= \delta_{nm} .\label{fermions_anticommute_prop_2}
\end{align}
Fermion operators $a_n^\dagger$ and $a_n$ can be equally written in terms of the Majorana operators $\gamma_{(n, l)}$ and $\gamma_{(n, r)}$ as
\begin{align}
 a_n^\dagger &= \frac{1}{2}(\gamma_{(n, l)} - i\gamma_{(n, r)}) \label{fermion_operator_dagger}\\
 a_n &= \frac{1}{2}(\gamma_{(n, l)} + i\gamma_{(n, r)}) .\label{fermion_operator}
\end{align}
Using simple algebra we can demonstrate that we can also easily transform them the other way around as follows
\begin{align}
 \gamma_{(n, l)} &= a_n + a_n^\dagger \label{majorana_operator_l}\\
 \gamma_{(n, r)} &= -ia_n + ia_n^\dagger. \label{majorana_operator_r}
\end{align}
Each Majorana operator is described by an index $(n,\sigma)$ composed of two values $n$, the 1D lattice site index, and $\sigma$ the Majorana species where $\sigma \in \{l, r\}$. The $l$ and $r$ labels denote the ``left'' and ``right'' Majorana species. Thus each fermion can be represented by box with two compartments, as shown in Fig. \ref{trivial_topological_pairings}(a). A Majorana can occupy or not occupy each box.  While the $l$ and $r$ Majoranas correspond to the left and right boxes, we emphasize that these are nothing to do with spatial degrees of freedom (much like spin ``up'' and ``down'' are not spatial). Operators for MZMs also satisfy
\begin{align}
 \gamma_k &= \gamma_k^\dagger, \label{fermions_anticommute_prop_2}\\
 \{\gamma_{(n, \sigma)}, \gamma_{(n^\prime, \sigma^\prime)}\} &= 2\delta_{nn^\prime} \delta_{\sigma\sigma^\prime}, \label{majorana_fermions_anticommute}
\end{align}
where (\ref{fermions_anticommute_prop_2}) can be interpreted as Majorana fermion being its own antiparticle. Since Majoranas are fermions any two distinct MZMs must anticommute (\ref{majorana_fermions_anticommute}).

The transformation (\ref{fermion_operator_dagger}) and (\ref{fermion_operator}) can be considered merely a change of variables between a bare fermion and a pair of underlying MZMs.  A more non-trivial transformation results by taking two MZMs from non-neighboring pairs $\alpha=(n, \sigma)$, $\beta=(m, \nu)$ and constructing a new kind of fermion $f_{\alpha\beta}^\dagger$ and $f_{\alpha\beta}$, corresponding to 
\begin{align}
 f_{\alpha\beta}^\dagger &= \frac{1}{2}(\gamma_\alpha - i\gamma_\beta), \label{quasifermion_operator_dagger}\\
 f_{\alpha\beta} &= \frac{1}{2}(\gamma_\alpha + i\gamma_\beta), \label{quasifermion_operator}
\end{align}
From now on we will refer to the delocalized fermions as ``quasifermions''. For the localized fermions, we will simply refer to them as regular fermions. When we write fermions we mean a general particle which satisfies the fermionic properties without specifying if it is localized or not. We can formally state it by defining pairs $\alpha\beta = (n, \sigma)(m, \nu)$ and $\alpha^\prime \beta^\prime = (n^\prime, \sigma^\prime)(m^\prime, \nu^\prime)$ and then the anti-commutation relations can be evaluated as
\begin{align}
 \{f_{\alpha\beta}, f_{\alpha^\prime \beta^\prime}\} &= \frac{1}{2}\delta_{\alpha\alpha^\prime} - \frac{1}{2}\delta_{\beta\beta^\prime} +i\frac{1}{2}\delta_{\alpha\beta^\prime} +i\frac{1}{2}\delta_{\beta\alpha^\prime} \label{quasifermion_anticommutation_1} \\
 \{f_{\alpha\beta}, f_{\alpha^\prime \beta^\prime}^\dagger\} &= \frac{1}{2}\delta_{\alpha\alpha^\prime} + \frac{1}{2}\delta_{\beta\beta^\prime} -i\frac{1}{2}\delta_{\alpha\beta^\prime} +i\frac{1}{2}\delta_{\beta\alpha^\prime}. \label{quasifermion_anticommutation_2}
\end{align}
Here we introduce an assumption that any two valid quasifermions $f_{\alpha\beta}$ and $f_{\alpha^\prime \beta^\prime}$ have no Majoranas in common, i.e. $\{\alpha, \beta\} \cap \{\alpha^\prime, \beta^\prime\} = \emptyset$. Under this non-overlapping assumption commutation relations (\ref{quasifermion_anticommutation_1}) and (\ref{quasifermion_anticommutation_2}) reduce to
\begin{align}
 \{f_{\alpha\beta}, f_{\alpha^\prime \beta^\prime}\} &= 0, \label{quasifermion_anticommutation_revised_1} \\
 \{f_{\alpha\beta}, f_{\alpha^\prime \beta^\prime}^\dagger\} &= \delta_{\alpha\alpha^\prime}\delta_{\beta\beta^\prime}, \label{quasifermion_anticommutation_revised_2}
\end{align}
making the quasifermions equivalent to regular fermions. Using (\ref{quasifermion_operator_dagger})-(\ref{quasifermion_operator}) we can derive the phase factor exchange of Majoranas in such quasifermions
\begin{align}
 f_{\alpha\beta} &= -i f_{\beta\alpha}^\dagger.
\end{align}

\subsection{Nanowire Hamitonian}

Up to this point we have not been specific about which physical system to realize the 1DTS.  Some examples of systems that can realize 1DTS include nanowire semiconductor-superconductor heterostructures \cite{lutchyn2010majorana, oreg2010helical, mourik2012signatures, vanHeck2016, Zhang2018} or carbon nanotubes \cite{SanJose2015, Marganska2018, Lesser_2020}. For the sake of concreteness we henceforth consider semiconductor nanowires, and describe physical quantities in reference to this system.  We emphasize that this is for readability and the formalism should be equally applicable to any 1DTS system.  

We model the semiconductor nanowire using the Kitaev chain \cite{kitaev2001unpaired}, which has the following form
\begin{align}
 H = -&\sum^M_{j=1} \mu_j a_j^\dagger a_j - \sum^{M-1}_{j=1} t_j ( a_{j+1}^\dagger a_j + a_j^\dagger a_{j+1} )\nonumber \\
 + &\sum^{M-1}_{j=1} \Delta_j (a_j a_{j+1} + a_{j+1}^\dagger a^\dagger_j). \label{hamiltonian_fermionic_pre}
\end{align}
The above Hamiltonian possesses different phases depending on the choice of parameters: $\mu_j$ is the on-site energy on site $j$, $t_j$ is the coefficient of the electron hopping terms between neighboring lattice sites $j$ and $j+1$, and $\Delta_j$ is the BCS coupling describing Cooper pairing between the sites $j$ and $j+1$.

For our purposes we will only consider the regime where $\Delta_j = -t_j$ and reduce the number of parameters leading to a slightly simpler form \cite{Backens2017}
\begin{align}
 H = -&\sum^M_{j=1} \mu_j a_j^\dagger a_j \nonumber \\
 - &\sum^{M-1}_{j=1} t_j ( a_{j+1}^\dagger a_j + a_j^\dagger a_{j+1} + a_j a_{j+1} + a_{j+1}^\dagger a^\dagger_j). \label{hamiltonian_fermionic}
\end{align}
We can use the variable change described by (\ref{fermion_operator_dagger}) and (\ref{fermion_operator}) to rewrite the Hamiltonian (\ref{hamiltonian_fermionic}) in terms of $\gamma_{(n, l)}$ and $\gamma_{(n, r)}$ operators
\begin{align}
 H = -&\frac{1}{2}\sum^M_{j=1} \mu_j (1 + i\gamma_{(j,l)} \gamma_{(j, r)})
 - i \sum^{M-1}_{j=1} t_j\gamma_{(j,r)}\gamma_{(j+1,l)}.\label{hamiltonian_majorana_full}
\end{align}
We assume the parameters $\mu_j$ and $t_j$ from (\ref{hamiltonian_fermionic}) and (\ref{hamiltonian_majorana_full}) can be individually tuned. Specifically, the chemical potential $ \mu_j $ is controlled by the keyboard gates and the couplings $ t_j $ can be broken in particular places by the junction gates in Fig. \ref{processor_design}.

\subsection{Phases of the nanowire system}

There are two important physical phases of the Hamiltonian (\ref{hamiltonian_fermionic}) --- a topologically trivial regime which we will also refer to as \textit{normal phase} (or $N$-phase) and the \textit{topological regime} (or $T$-phase). The $T$-phase is present under condition $\mu < 2 \vert t_j \vert$ and otherwise the system enters the $N$-phase \cite{kitaev2001unpaired}. To better illustrate the difference between these phases we will focus on the limiting cases of those parameters where it is possible to write a simple expression for the eigenspectrum of the Hamiltonian (\ref{hamiltonian_fermionic}). The limiting cases that we consider are $\mu_j > 0, t_j = 0$ for the $N$-phase and $\mu_j = 0, t_j > 0$ for the $T$-phase. While we consider these limiting cases for simplicity in this section, the operations that we consider in this work are effective as long as the system remains in the required physical phase.

\subsubsection{Normal phase}

For the parameters $\mu_j > 0, t_j = 0$ nanowire reaches the limiting case of the $N$-phase characterized by the on-site pairing between the Majoranas, meaning the paired Majorana fermions lay on the lattice sites. Such pairing is the default pairing most commonly occurring in non-superconducting systems. A Hamiltonian term representing the $j$th on-site pairing is
\begin{align}
 H^{(N)}_j &= 1 + i \gamma_{(j, l)} \gamma_{(j, r)}. \label{hamiltonian_majorana_trivial_terms}
\end{align}
A complete $N$-phase Hamiltonian can be decomposed into a sum of terms of the form (\ref{hamiltonian_majorana_trivial_terms}) parametrized by their corresponding on-site energy $\mu_j$ values
\begin{align}
 H^{(N)} &= -\sum^M_{j=1} \mu_j a_j^\dagger a_j \label{hamiltonian_majorana_trivial_fermionic} \\
&= -\sum^M_{j=1} \mu_j f_{(j,l)(j,r)}^\dagger f_{(j,l)(j,r)} \label{hamiltonian_majorana_trivial_quasifermionic} \\
&= -\frac{1}{2}\sum^M_{j=1} \mu_j (1 + i\gamma_{(j,l)} \gamma_{(j, r)}) \label{hamiltonian_majorana_trivial} \\
&= -\sum^M_{j=1} \mu_j H^{(N)}_j. \label{hamiltonian_majorana_trivial_terms}
\end{align}
Let us label the eigenstates of (\ref{hamiltonian_majorana_trivial}) as all the possible placements of fermions in the nanowire. We denote $\ket{0}_N$ to be the $N$-phase vacuum state defined as the state such that $a_n \ket{0}_N = 0$. An arbitrary eigenstate can be written as
\begin{align}
 \ket{l_1\dots l_{M-1} l_M}_N &= \prod_n (f^\dag_{(n,l)(n,r)})^{l_n} \ket{0}_N \\
 &= \prod_n (a^\dag_n)^{l_n} \ket{0}_N.
\end{align}
Here the presence or absence of fermion at particular site $n$ is indicated by value $l_n \in \{0, 1\}$.

The energy spectrum of $H^{(N)}$ is
\begin{align}
 E_N &= \ev{H^{(N)}}{l_1\dots l_{M-1} l_M}_N \nonumber \\
 &= \sum^{M}_{j=1} \mu_j l_j, \label{energy_spectrum_topologically_trivial_regime}
\end{align}
which falls in the range $0 \leq E_N \leq M\mu$ and is characterized by the number of fermions in the nanowire. The types of fermions that build this kind of spectrum have a Majorana pairing as in Fig. \ref{trivial_topological_pairings}(a). Under this regime the paired Majorana fermions correspond to regular fermions.

\subsubsection{Topological phase}

\begin{figure}[t!]
    \includegraphics[height=0.5\linewidth]{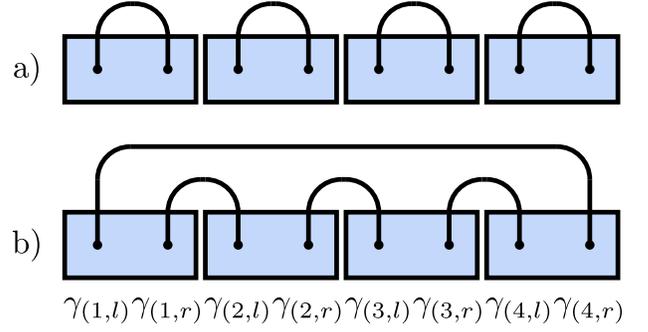}
    \caption{(a) the normal phase $N $, (b) the topological phase $T$.}
    \label{trivial_topological_pairings}
\end{figure}

The $T$-phase is a phase in which the quasifermions underlying Majorana pairs originate from different sites. The $j$th inter-site pairing term is defined as
\begin{align}
 H^{(T)}_j &= i \gamma_{(j, r)}\gamma_{(j+1,l)} \label{hamiltonian_majorana_topological_terms_definition}
\end{align}
and in the limiting case $\mu_j = 0, t_j > 0$, the Hamiltonian (\ref{hamiltonian_majorana_full}) can be decomposed into a sum of terms (\ref{hamiltonian_majorana_topological_terms}) parametrized by their corresponding electron hopping $t_j$ values
\begin{align}
 H^{(T)} &= -2 \sum^{M-1}_{j=1} t_j f^\dagger_{(j,r)(j+1,l)} f_{(j,r)(j+1,l)} \label{hamiltonian_majorana_topological_quasifermionic} \\
 &= i \sum^{M-1}_{j=1} t_j \gamma_{(j, r)}\gamma_{(j+1,l)} \label{hamiltonian_majorana_topological_plus} \\
 &= \sum^{M-1}_{j=1} t_j H^{(T)}_j. \label{hamiltonian_majorana_topological_terms}
\end{align}
The above Hamiltonian takes the form of a sum of terms representing a fermionic number operator capable of detecting particular pairing of Majoranas by taking the expectation value of
\begin{align}
  {\cal N}_{\alpha\beta} &= f_{\alpha\beta}^\dagger f_{\alpha\beta} \label{number_operator}.
\end{align}
The $T$-phase Hamiltonian has an off-site pairing between sites and leaves two remaining Majoranas paired between the first and last site on the outer edges as shown in Fig. \ref{trivial_topological_pairings}(b). To indicate the topological pairing of the Majoranas, we use a $T$ superscript on the Hamiltonian. The form (\ref{hamiltonian_majorana_topological_quasifermionic}) is obtained from (\ref{hamiltonian_majorana_topological_plus}) by applying the inverse transformation (\ref{quasifermion_operator_dagger})-(\ref{quasifermion_operator}). The diagonal form (\ref{hamiltonian_majorana_topological_quasifermionic}) is more suitable for studying the energy spectrum of the $T$-phase. The $T$-phase eigenstates are built from the $T$-phase vacuum state
\begin{align}
  f_{(n,r)(n+1,l)} \ket{0}_{T} &= 0 \\ 
  f_{(1,l)(M,r)} \ket{0}_{T} &= 0 
\end{align}
and are defined as
\begin{align}
  \ket{l_1\dots l_{M-1} l_M}_{T}
 =& (f^\dag_{(1,l)(M,r)})^{l_M} \prod_{1 \leq j < M} (f^\dag_{(j,r)(j+1,l)})^{l_j} \ket{0}_{T}. \label{topological_state}
\end{align}
The energy spectrum of the $T$-phase Hamiltonian is
\begin{align}
 E_T &= \ev{H^{(T)}}{l_1\dots l_{M-1} l_M}_T \nonumber \\
 &= -2\sum^{M-1}_{j=1} t_{j} l_j. \label{energy_spectrum_topological_regime}
\end{align}

The $T$-phase states are characterized by a quasifermion pairing that is highly delocalized between the furthermost lattice sites. The occupation of this highly delocalized quasifermion is by convention represented by the last index $l_M$. To visualize this we illustrate the pairing of the quasifermions in Fig. \ref{trivial_topological_pairings}(b). The highly delocalized quasifermion is composed of the zero energy Majoranas as can be seen from (\ref{energy_spectrum_topological_regime}), and are henceforth called the Majorana Zero Modes (MZMs). The MZMs do not contribute to the energy level of the nanowire system and can be thought to have a zero coefficient on the delocalized pairing terms in (\ref{energy_spectrum_topological_regime}). The remaining indices represent the off-site pairing as seen in Fig. \ref{trivial_topological_pairings}(b). Topological states and normal phase states are eigenstates of different Hamiltonians and they are inherently not compatible. Thus the operator $a_n^\dagger$ cannot be used to create eigenstates of the Hamiltonian $H^{(N)}$ by application on $\ket{0}_{T}$, for example.

\subsection{Braiding operators}

\begin{figure}[t!]
    \includegraphics[height=0.5\linewidth]{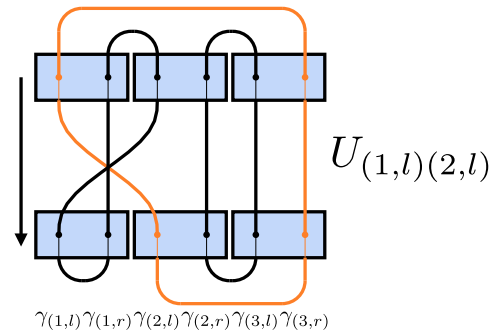}
    \caption{Visual notation in which rows are topological quantum states characterized by a particular pairing and transitions between rows are braids. On this example we can see the effect of $U_{(1,l)(2,l)}\ket{001}_{T} = \ket{0}_N \otimes \ket{01}_{T}$.}
    \label{figure_braiding_example}
\end{figure}

Changing the arrangements of Majoranas is achieved using a braiding operator \cite{Ivanov2001, Nayak1996}
\begin{align}
  U_{\alpha\beta} &= \exp{\frac{\pi}{4}\gamma_\alpha \gamma_\beta} \label{braiding_operator} \\
  &= \frac{1}{\sqrt{2}}(1 + \gamma_\alpha \gamma_\beta). \nonumber
\end{align}
The above operator has the effect of changing a Majorana fermion into another Majorana fermion
\begin{align}
  U_{\alpha\beta} \gamma_\alpha U_{\alpha\beta}^\dagger &= -\gamma_\beta \label{braiding_operator_identity_1} \\
  U_{\alpha\beta} \gamma_\beta U_{\alpha\beta}^\dagger &= \gamma_\alpha. \label{braiding_operator_identity_2}
\end{align}
An example of such interchange between the two Majorana modes $\gamma_{(2,r)}$ and $\gamma_{(3,r)}$ is shown in Fig. \ref{figure_braiding_example}. The swapping effect of braiding can be summarized by showing how the braiding operator acts on the annihilation operator of a quasifermion
\begin{align}
  U_{\alpha\beta} f_{\alpha\beta} U_{\alpha\beta}^\dagger &= i f_{\beta\alpha} \label{braiding_operator_identity_3} \\
  U_{\beta\alpha} f_{\alpha\beta}^\dagger  U_{\beta\alpha}^\dagger &= -i f_{\beta\alpha}^\dagger. \label{braiding_operator_identity_4}
\end{align}
Finding a particular sequence of swaps can lead to changing one topological configuration into another. This is how quantum information is processed using MZMs.

\subsection{Conversion of ground states by braiding}
\label{section_conversion_ground_states_braiding}

One way to change the state between the two phases is to perform an adiabatic variation of the parameters of the Hamiltonian, such that the phase transition $\mu_j = 2 t_j$ is crossed. An alternative way is to directly apply braiding operations to the ground state to convert the normal state into a topological state and vice versa. Being able to convert $N$-phase into $T$-phase is important because those phases are inherently incompatible in the way that $a_j^\dagger$ cannot act on $\ket{0\dots0}_T$ and $f_{nm}^\dagger$ cannot act on $\ket{0\dots0}_N$ to produce an eigenstate of the corresponding Hamiltonians.

\begin{figure}[t]
\includegraphics[width=\linewidth]{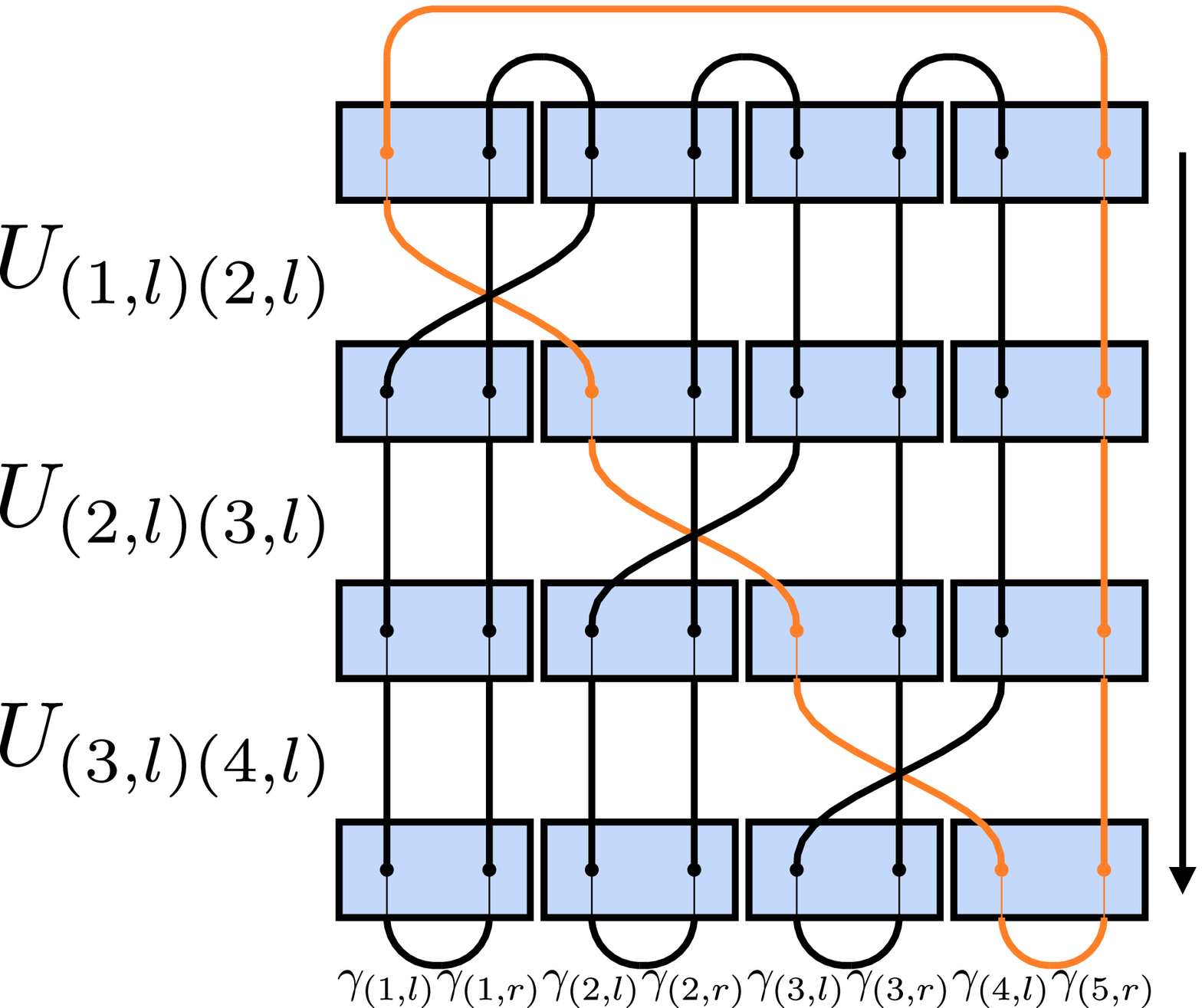}
\caption{A step by step procedure of converting $T$-phase state into $N$-phase state by conversion operator $U_c$.}\label{uc_example}
\end{figure}

A braiding sequence turning the $N$-phase into the $T$-phase can be achieved by applying the conversion operator
\begin{align}
  U_c &= U_{(1,l)(2,l)} U_{(2,l)(3,l)} \dots U_{(M-1,l)(M,l)} \nonumber \\
  &= \prod_{j=1}^{M-1} U_{(j,l)(j+1,l)}. \label{conversion_operator}
\end{align}
A visualization of how a sequence of local braids equivalent to $U_c$ can turn the $T$-phase into $N$-phase and vice versa is provided in Fig. \ref{uc_example}. The product notation in (\ref{conversion_operator}) is ambiguous so we assume the convention that the product should be always expanded from left to right, in order from the lower index to the upper index of the product sign. Such a product of braiding operations is a sequence of braids which can be visually represented by a sequence of swapping of paired lattice sites (Fig. \ref{figure_braiding_example}). First note that $U_c$ only involves braiding of $l$-type Majoranas, hence the $r$ Majoranas are unaffected 
\begin{align}
    U_c \gamma_{(n,r)} U_c^\dagger &= \gamma_{(n,r)}.
\end{align}
Now with exception of the $ \gamma_{L,l} $ Majorana, the effect of $U_c$ is to shift to one site to the right of the existing site
\begin{align}
    U_c \gamma_{(n,l)} U_c^\dagger &= -\gamma_{(n+1,l)}.
\end{align}
The exception to this is the $l$-Majorana on the right-most site, which gets transported all the way to the left-most site
\begin{align}
    U_c \gamma_{(M,l)} U_c^\dagger &= \gamma_{(1,l)}.
\end{align}
Using these relations, one may deduce the effect of $U_c$ on the fermions as 
\begin{align}
  U_c a_n U_c^\dagger =
  \begin{cases} - f_{(n, l)(n+1, l)} &\mbox{if } n < M \\
  f_{(1, l)(M, r)} & \mbox{otherwise.} \end{cases} \label{converting_fermions}
\end{align}
Therefore acting $U_c$ on a normal state induces a phase transition delocalizing the rightmost fermion placed at index $n=M$ into quasifermion delocalized between the first and last sites of the chain
\begin{align}
  U_c H^{(N)} U_c^\dagger =& U_c(-\sum^M_{j=1} \mu_j a_j^\dagger a_j)U_c^\dagger. \label{spectrum_mapping_2} \\
  =& -\sum^M_{j=1} \mu_j U_c a_j^\dagger U_c^\dagger U_c a_j U_c^\dagger. \label{spectrum_mapping_3} \\
  =& -\mu_M f_{(1,l)(M,r)}^\dagger f_{(1,l)(M,r)} \nonumber \\
  &-\sum^{M-1}_{j=1} \mu_j f_{(j,l)(j+1,l)}^\dagger f_{(j,l)(j+1,l)} \label{spectrum_mapping_5}
\end{align}
Expression (\ref{spectrum_mapping_3}) is obtained by injecting the identity $I = U_c^\dagger U_c$ in the middle. Next, we apply (\ref{converting_fermions}) to convert fermions into quasifermions. We recognize the resulting term (\ref{spectrum_mapping_5}) as equivalent to the $T$-phase Hamiltonian (\ref{hamiltonian_majorana_topological_plus}) up to transformation $t_j \rightarrow \mu_j$. Therefore (\ref{spectrum_mapping_5}) shares the same eigenstates as the $T$-phase Hamiltonian
\begin{align}
  U_c H^{(N)} U_c^\dagger \ket{l_1 \dots l_M}_T = E_N^c \ket{l_1 \dots l_M}_T
\end{align}
where
\begin{align}
  E_N^c = -\mu_M l_M - \sum_{j=1}^{M-1} \mu_j l_j.
\end{align}
The rightmost site energy $\mu_M$ is the MZM's energy, which in the topological Hamiltonian $H^{(T)}$ is zero. Our convention is that its the rightmost site that is being delocalized into a quasifermion representing MZMs.

Ground states of the $N$-phase and $T$-phase are ground states of different Hamiltonians and thus are not compatible, also making their operators incompatible, in the general case. However since we have the $U_c$ conversion operator, we can use it to cause a phase transition between those phases
\begin{align}
  U_c \ket{0}_N &\propto \ket{0}_{T} \label{uc_effect}
\end{align}
which makes it possible to convert delocalized fermions into regular fermions, process them and delocalize again. The above expression is equal up to global phase. We will use this idea later to develop quantum gates.

\subsection{Logical states}

The logical states are defined by the pairing of MZMs on the domain edges
\begin{align}
  \ket{0_L} &= \ket{0}_{T} \label{logic_0} \\
  \ket{1_L} &= f_{(1,l)(M,r)}^\dagger \ket{0}_{T}. \label{logic_1}
\end{align}
As the logic states are eigenstates of (\ref{hamiltonian_majorana_topological_quasifermionic})-(\ref{hamiltonian_majorana_topological_terms}) from now we assume the system is in the $T$-phase. If more than a single logical qubit is involved in the computation these are stored on the separate nanowires. An example of a topological quantum gate on such qubit is the logical $\sqrt{Z}$-gate operation which is achieved by performing a braid of the MZMs on the same chain 
\begin{align}
  U_{(1,l)(M,r)} &= \sqrt{Z} \label{gate_z1_single}
\end{align}
which follows from
\begin{align}
  U_{(1,l)(M,r)} f_{(1,l)(M,r)} U_{(1,l)(M,r)}^\dagger &= i f_{(1,l)(M,r)} \label{gate_z1_single_deriv_1} \\
  U_{(1,l)(M,r)} f^\dagger_{(1,l)(M,r)} U_{(1,l)(M,r)}^\dagger &= -i f^\dagger_{(1,l)(M,r)}. \label{gate_z1_single_deriv_1}
\end{align}
One may easily verify that the appropriate phase factor corresponding to a $Z$-operation is realized for the logical state $\ket{0_L}$ and $\ket{1_L}$.

Suppose that the only operations that can be performed are to exchange the MZMs on the edge of each domain, using the braiding operators (\ref{braiding_operator}).  In this case, if we assume two chains, there are only six possible gates that can be implemented \cite{hassler2014majorana}. This follows from the fact that for two domains there are four edges and it leads to six possible pairings and thus six possible braids. In terms of the logical space operations those six gates can be summarized as follows \cite{hassler2014majorana}
\begin{align}
  U_{l_1 r_1} &= \sqrt{Z_1} \label{gate_z1} \\
  U_{l_1 l_2} &= \sqrt{Y_1 X_2} \label{gate_y1x2} \\
  U_{l_1 r_2} &= \sqrt{Y_1 Y_2} \label{gate_y1y2} \\
  U_{r_1 l_2} &= \sqrt{X_1 X_2} \label{gate_x1x2} \\
  U_{r_1 r_2} &= \sqrt{X_1 Y_2} \label{gate_x1y2} \\
  U_{l_2 r_2} &= \sqrt{Z_2}. \label{gate_z2}
\end{align}
Where we indicate the left edge of first topological qubit to be $\gamma_{l_1}$ and its right edge to be $\gamma_{r_1}$ and the edges of the second topological qubit are $\gamma_{l_2}$ and $\gamma_{r_2}$.

\section{Logical $Z$ operation}

We now examine the protocol of Ref. \cite{Backens2017} and show how it is possible to produce an effective braiding gate with the addition of an auxiliary qubit.  This allows for a 
way of producing braiding gates without fabricating T-junctions. In Ref. \cite{Backens2017} an adiabatic sequence was used, but we shall rederive the protocol in a unitary language.  This allows not only for a clearer description of the scheme, but allows us to generalize the scheme to multiple logical qubit gates.  

\subsection{Local double braid sequence}

We first describe a scheme where the logical $Z$ operation is performed by a sequence of local braids that are applied in sequence through the lattice. While this is obviously more complex than directly performing a braid between only the MZMs such as in (\ref{gate_z1})-(\ref{gate_z2}), this will elucidate an equivalent scheme shown in the next section, where an auxiliary spin can achieve the same effect. This will be the basis for the full $Z$-rotation scheme described later. Since the MZMs are on either end of the topological domains, these braiding operations require non-local operations.

Equations (\ref{gate_z1})-(\ref{gate_z2}) showed examples of logical operations that can be performed by directly performing braiding operations on the MZMs.  Since the MZMs are on either end of the topological domains, these braiding operations require a non-local operations.  Such an operation is not simple to perform in a purely one-dimensional geometry as we consider in Fig. \ref{processor_design}.  The non-local nature of the operation is part of why the encoded information is resilent under decoherence, since such operations do not happen easily naturally.  

Here we introduce a sequence of completely local braiding operations which produces a logical $Z$ operation. The braiding sequence is applied to the entire chain, hence is still consistent with the notion that a topological operation is required to perform a logical operation.  However, each operation is a local operation, and occurs in a specified ordering, which makes it more accessible to a realistic gate operation.  

Let us consider the operation
\begin{align}
  U_Z &= \prod^{M}_{j=1}U_{(j,l)(j,r)}^2. \label{unitary_z_braiding}
\end{align}
Writing $U_Z$ in terms of Majorana operators we have
\begin{align}
  & U_Z = \prod^{M}_{j=1}\gamma_{(j,l)}\gamma_{(j,r)}.
\end{align}
The logical space is defined in terms of MZMs, isolating them from remaining Majoranas we have
\begin{align}
  U_Z =& \gamma_{(1,l)} (\prod^{M-1}_{j=2}\gamma_{(j,r)}\gamma_{(j+1,l)}) \gamma_{(M,r)} \\
  =& \gamma_{(1,l)} \gamma_{(M,r)} \prod^{M-1}_{j=2}\gamma_{(j,r)}\gamma_{(j+1,l)}.
\end{align}
The above can be rewritten using quasifermion number operators
\begin{align}
  U_Z =& i(1-2{\cal N}_{(1,l)(M,r)}) \prod^{M-1}_{j=2}i(1 - 2 {\cal N}_{(j,r)(j+1,l)}). \label{Uz_deriv_final} 
\end{align}
Note that each of the terms of this product gives a factor of $\pm 1 $ for a quasifermion number state. Applying (\ref{Uz_deriv_final}) to an arbitrary topological state of the form (\ref{topological_state}) gives
\begin{align}
 &U_Z \ket{l_1\dots l_{M-1} l_M}_{T} \\
 =& i^{M-1} (-1)^{l_{M}} \prod^{M-1}_{j=2} (-1)^{l_j} \ket{l_1\dots l_{M-1} l_M}_{T}. \label{uz_deriv_topological_state}
\end{align}
The logical space as defined (\ref{logic_0})-(\ref{logic_1}) is characterized by $l_k = 0$ for $2 \leq k \leq M-1$. This is a special case of (\ref{uz_deriv_topological_state}) which gives
\begin{align}
 U_Z \ket{0\dots0 l_M}_{T} =& i^{M-1} (-1)^{l_M} \ket{0\dots0 l_M}_{T}. \label{uz_deriv_logical_space_state}
\end{align}
Logical space states differ only by the configuration of MZMs, i.e. the value of $l_M$. It only affects the sign of the overall expression pulling out the eigenstates equivalent to $Z$.
\begin{align}
  U_Z \ket{0_L} &= i^{M-1}\ket{0_L} \\
  U_Z \ket{1_L} &= -i^{M-1}\ket{1_L}.
\end{align}

The sequence of braids (\ref{unitary_z_braiding}) is equivalent to the protocol in Ref. \cite{Backens2017} because every time the topological domain moves through the coupler a bit flip operation is applied to the site which is adjacent to the coupler after the move. Generalizing this operation and transforming back from the spin-$\frac{1}{2}$ language to the fermion language leads to the $U_{(j,l)(j,r)}^2$ operators applied sequentially.

We claim the $U_Z$ gate implemented in this way is topologically protected because the overall distance between the MZMs --- $M$ is not affected during the braid. The entire protocol consists only of local interactions within the nanowire ensuring the energy gap does not close. Another possible way of seeing this is during every single step of $U_Z$ the system remains an eigenstate of $T$-phase Hamiltonian, (\ref{hamiltonian_majorana_topological_plus}).

\subsection{Effect of $U_Z$ on the Hamiltonian}

The previous section showed how to perform a logical $Z$ operation using a sequence of local braids.  Our final aim for this section is to perform this gate entirely using adiabatic operations.  To this end, we deduce in this section the effect of the local braids on the Hamiltonian. As a starting point let us use the Hamiltonian (\ref{hamiltonian_majorana_topological_plus}) and transform it under $U_Z$, according to $U_Z H U_Z^\dagger$. To evaluate this, first let us find the effect of $U_{(j,l)(j,r)}^2 = \gamma_{(j,l)}\gamma_{(j,r)}$ on one of the terms within the $T$-phase Hamiltonian (\ref{hamiltonian_majorana_topological_terms_definition}). Let us start with the effect of the double braid operator on the Majoranas of both species $l$ and $r$
\begin{align}
U^2_{(j,l)(j,r)} \gamma_{(j,l)} (U^2_{(j,l)(j,r)})^\dagger &=  - \gamma_{(j,l)} \\
U^2_{(j,l)(j,r)} \gamma_{(j,r)} (U^2_{(j,l)(j,r)})^\dagger &=  - \gamma_{(j,r)}. 
\end{align}
We see that the double braid preserves the species of Majorana and only puts a minus sign on it. Then this naturally leads onto the result that 
\begin{align}
U^2_{(j,l)(j,r)} H^{(T)}_j (U^2_{(j,l)(j,r)})^\dagger = -H^{(T)}_j
\end{align}
since it contains only one Majorana on index $j$.

Applying $U_Z$ operator to $H^{(T)}_j$ leaves all the terms invariant as it interchanges the sign of $H_j$ twice, once each for the two adjacent sites on each term. We thus have
\begin{align}
U_Z H^{(T)}_j U_Z^\dagger = H^{(T)}_j,
\end{align}
and hence it follow that
\begin{align}
U_Z H^{(T)} U_Z^\dagger = H^{(T)},
\end{align}
thus the $T$-phase Hamiltonian (\ref{hamiltonian_majorana_topological_plus}) remains invariant under $U_Z$. When applied term by term, each double braid term $U_{(j,l)(j,r)}^2$ flips the sign of the $t_j$ parameter. This suggests a way of applying $U_Z$ adiabatically by changing the sign of $t_j$ twice consecutively throughout the chain. This is visually described in Fig. \ref{double_braid_moving}(a).

\subsection{Equivalent adiabatic scheme}

\begin{figure*}[t]
\begin{tabular}{cc}
  \includegraphics[height=8.5cm]{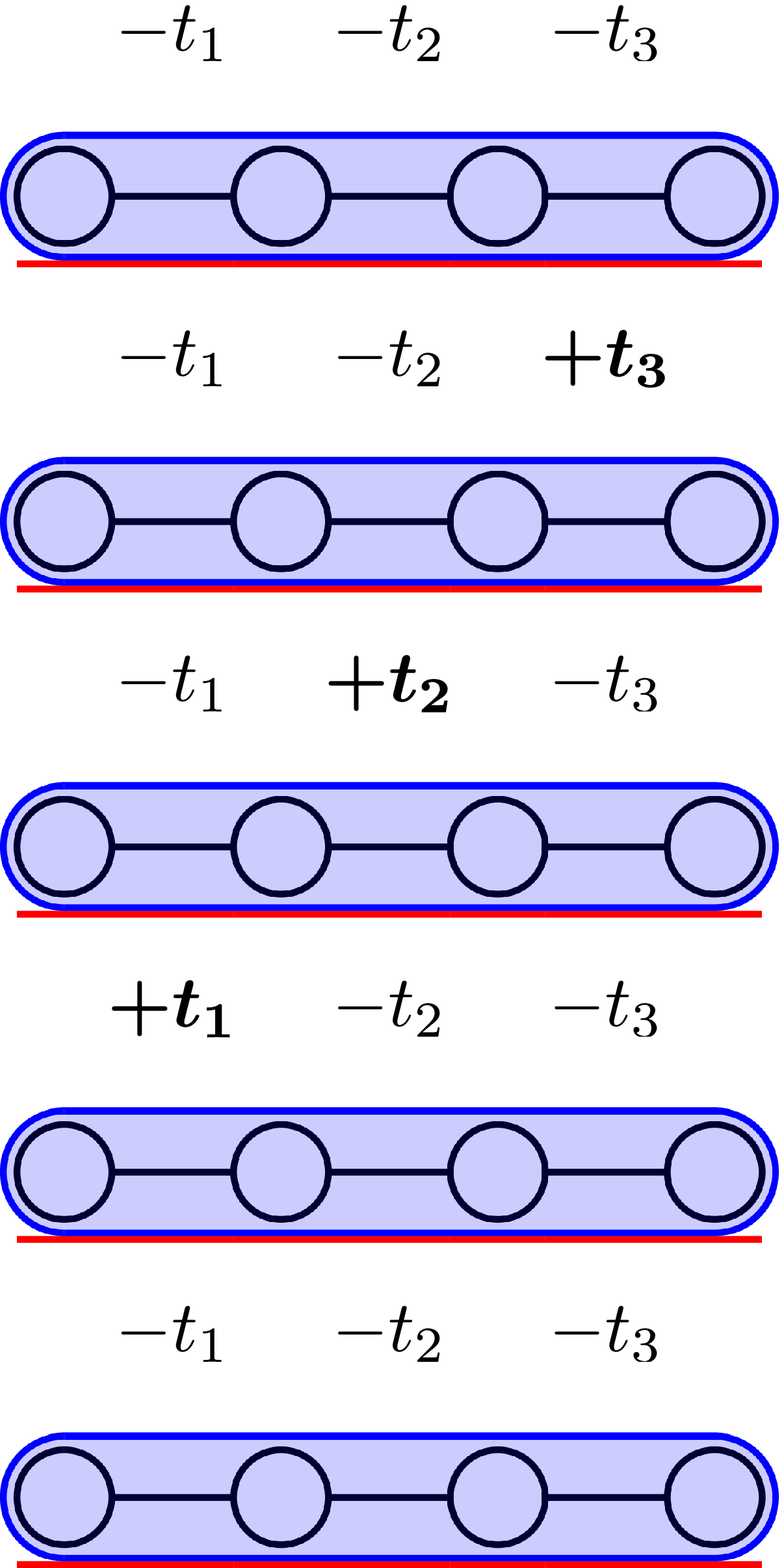} &
  \includegraphics[height=8.5cm]{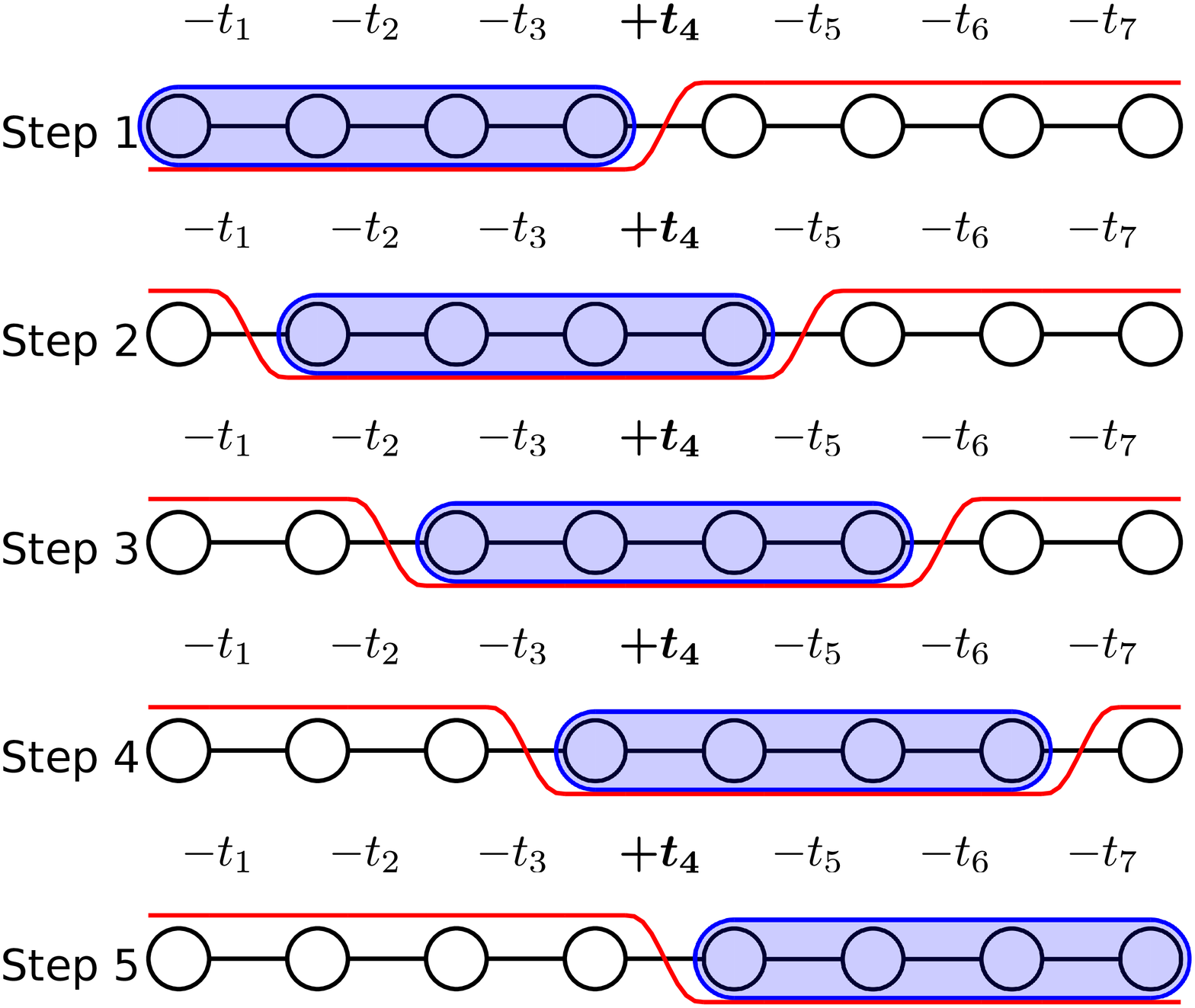} \\
(a) & (b) \\[6pt]
\end{tabular}
\caption{A diagram demonstrating how double braid operation is applied to a domain (a) by moving a domain through a region in which coupling parameter is reversed (b).}\label{double_braid_moving}
\end{figure*}

We now describe an equivalent adiabatic sequence for applying the $U_Z$ operator. The basic idea is shown in Fig. \ref{double_braid_moving}.  Since the effect of the sequence of $U^2_{(j,l)(j,r)}$ is equivalent to locally changing the sign of the $t_j$ (and $\Delta_j$, as we assumed earlier that $\Delta_j = t_j$), we consider a situation where there is a region of $+t_j$, adjacent to the chain. Then the chain is moved through the region of $+t_j$. By moving the chain through this $+t_j$ region, this is equivalent to applying the $U_Z$ sequence. We also note that the $+t_j$ has a reversed sign which by default was assumed to be $-t_j$.

In order to move the chain, we must first describe some elementary moves which can be combined to perform the whole sequence.  Specifically, we show the steps required for (i) moving a topological domain; (ii) performing a double braid while moving. 

\subsubsection{Moving topological domains}

\begin{figure}[t]
\includegraphics[width=\linewidth]{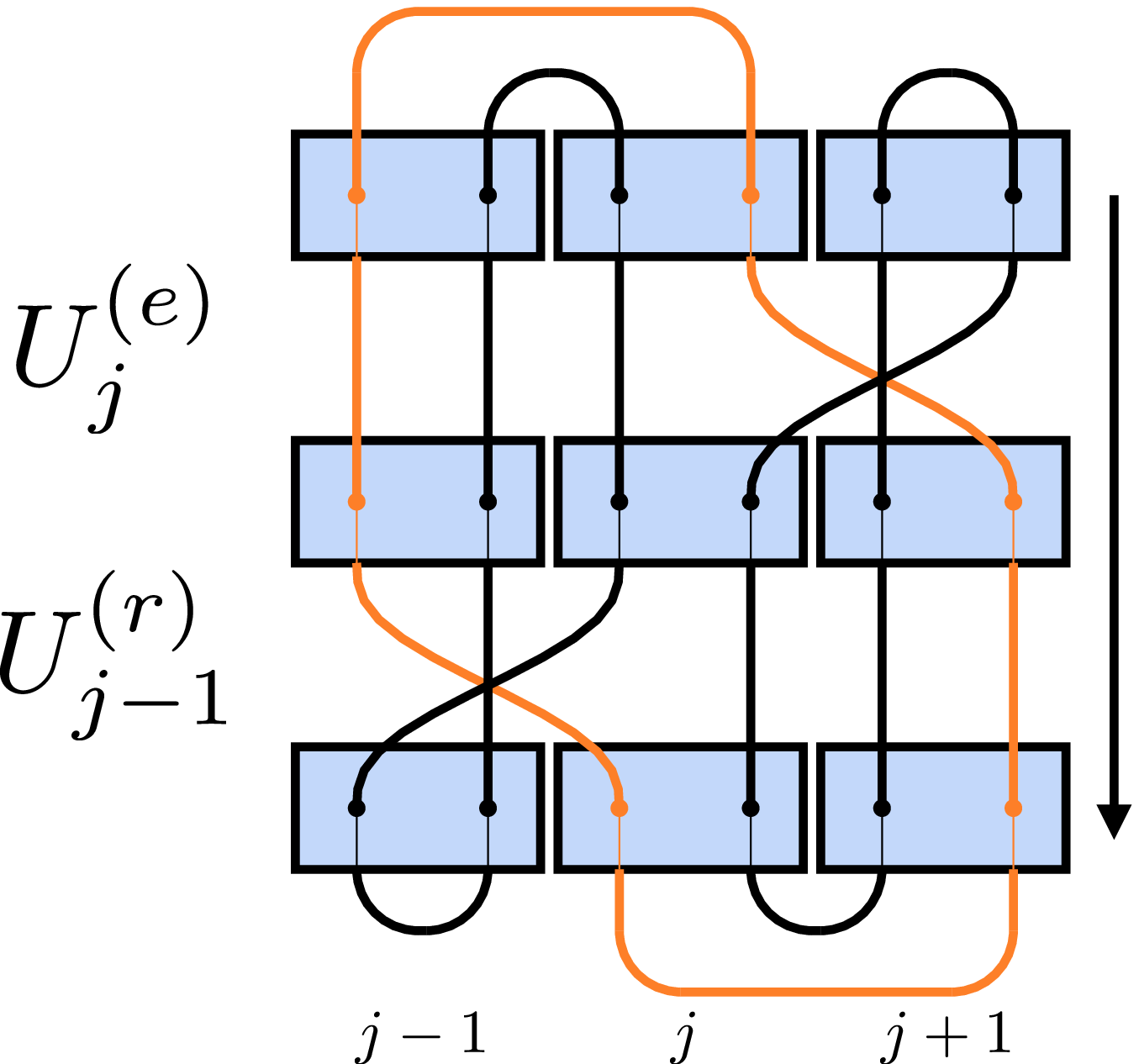}
\caption{A process of moving the domain by a single site to the right. The extension operator $U^{(e)}_{j}$ extends the domain by one site and then the retraction operator $U^{(r)}_{j-1}$ retracts the domain by one site from left to right. Overall effect is the moving operator $U^{(m)}_{(j-1)(j)} = U^{(r)}_{j-1}U^{(e)}_{j}$ which moves entire domain by one site.}\label{figure_moving_domain}
\end{figure}

In the sequence shown in Fig. \ref{double_braid_moving} (a), the chain in the active topological region is moved through a region of $+t_j$, which allows one to apply the $U_Z$ operation. In order to move the chain by one lattice site to the right, we need to create an extra site on the right side of the chain that is in the topological configuration, and remove one site on the left side and convert is to a normal phase region. We thus require operations for converting regions of the chain from a normal phase configuration to a topological phase configuration and vice-versa. The conversion operator $U_c$ (\ref{conversion_operator}) we introduced earlier converts performs a phase transition of an entire chain between $N$-phase and $T$-phase. By applying an equivalent approach of local braiding, we can derive operators which create a local phase transition in the vicinity of the ends of the topological domain. For that we introduce
\begin{align}
  U^{(e)}_{j} &= U_{(j+1, r)(j, r)} \label{operator_extension} \\
  U^{(r)}_{j} &= U_{(j, l)(j+1, l)} \label{operator_retraction} \\
  U^{(m)}_{jk} &= U^{(r)}_{j} U^{(e)}_{k} \label{operator_moving}
\end{align}
where $U^{(e)}_{j}$ is the extension operator, which extends the domain by a single site $j$ by making it join the $T$-phase on its left. The operator $U^{(r)}_{j}$ is the retraction operator which retracts the domain making the site $j$ leave the $T$-phase and become a $N$-phase site. The moving operator $U^{(m)}_{j,k}$ is the combination of the above two causing the overall effect of moving the domain of length $k-j$ by one site to the right. This effect is shown in Fig. \ref{figure_moving_domain}. The corresponding adiabatic protocol can be derived by studying the effect of those local phase transition braiding operators on the terms of the $N$-phase (\ref{hamiltonian_majorana_trivial_terms}) and $T$-phase (\ref{hamiltonian_majorana_topological_terms}) Hamiltonians. We examine the relevant case of those terms at the boundaries of the topological domain inside of the nanowire
\begin{align}
  U^{(e)}_{k} H^{(N)}_{k+1} (U^{(e)}_{k})^\dagger &= H^{(T)}_{k} \label{extend_n} \\
  U^{(e)}_{k} H^{(T)}_{k-1} (U^{(e)}_{l})^\dagger &= H^{(T)}_{k-1} \label{extend_t} \\
  U^{(r)}_{j} H^{(T)}_{j} (U^{(r)}_{j})^\dagger &= H^{(N)}_{j} \label{retract_tl} \\
  U^{(r)}_{j} H^{(T)}_{j+1} (U^{(r)}_{j})^\dagger &= H^{(T)}_{j+1} \label{retract_tr}
\end{align}
and we find how to manipulate the coefficients of (\ref{hamiltonian_majorana_full}) to implement operator (\ref{operator_moving}). From (\ref{extend_n}) we can deduce that to extend the right boundary of the domain from site $k$ to site $k+1$ according to (\ref{operator_extension}), we need to adiabatically reduce the local chemical potential $\mu_{k+1}$ and increase $t_k$. For retraction of the domain from site $j$ to site $j+1$ we adiabatically reduce the hopping $t_j$ and increase $\mu_j$, which achieves the equivalent of the $U^{(r)}_{j}$ operator.

We now show a simple example of the moving operator defined as above. Let us consider a nanowire of length $M$ encoding an arbitrary logical state
\begin{align}
\ket{\psi_{jk}} &= \ket{0_1 0_2\dots0_{j-1}}_N \nonumber \\
&\otimes (\alpha \ket{00\dots0}_{T} + \beta \ket{0\dots01}_{T}) \nonumber \\
&\otimes \ket{0_{k+1} 0_{k+2}\dots0_M}_N
\end{align}
where $\vert \alpha \vert^2 + \vert \beta \vert^2 = 1$. Here, there is a $T$-phase connecting sites $j$ and $k$ and remaining sites are in the $N$-phase. We assume $k < M$, then applying $U^{(m)}_j$ operator moves the domain by a single site
\begin{align}
U^{(m)}_{j,k} \ket{\psi_{jk}} &= \ket{\psi_{(j+1)(k+1)}}.
\end{align}
As described above, the same effect can be achieved by manipulating the coefficients of (\ref{hamiltonian_majorana_full}) Hamiltonian, by decreasing $\mu_{k+1}, t_j$ and increasing $\mu_j, t_k$. Hence, applying $U^{(m)}_j$ consists of changing the edges of the domain between topologically trivial and $T$-phase and the overall effect of it is shifting the domain to the right by one site. Applying it in the other way around, i.e increasing $\mu_{k+1}, t_j$ and decreasing $\mu_j, t_k$ results in shifting the domain to the left by one site.

\subsubsection{Performing an on-site braid while moving}
\label{section_performing_braid_while_moving}

As we explained in the previous section, manipulating the coefficients $\mu_j$ and $t_j$ can be used to move the topological domain inside of the nanowire. It has also been shown in Sec. III B that changing the sign of $t_j$ terms is equivalent to performing a double on-site braid. In this section we will explain how such a double on-site braid can be performed along the entire chain by moving it through a region that has reversed signs of $t_j$.

Let us consider a chain of total length $2M$. The doubled length is required to move the entire domain through the region of $+t_j$. The process of moving involves $M$ steps, at step $k=0$ the first $M$ sites are in the $T$-phase and last $M$ sites are in the $N$-phase
\begin{align}
  H_{(\text{step }0)} =& \sum_{j=1}^{M-1} t_j H^{(T)}_j + \sum_{j=M+1}^{2M} \mu_j H^{(N)}_j. \label{moving_0}
\end{align}
For steps $1 < k < M$, the $T$-phase is moved from the left to the right. Note that at site $k=M$, the sign of $t_j$ is reversed representing the special region, as shown in Fig. \ref{double_braid_moving}(b)
\begin{align}
  H_{(\text{step }k)} =& \sum_{j=1}^{k} \mu_j H^{(N)}_j + \sum_{\substack{j=k+1\\ j \neq M}}^{M+k} t_j H^{(T)}_j \nonumber \\
  - & t_M H^{(T)}_M + \sum_{j=M+k}^{2M} \mu_j H^{(N)}_j. \label{moving_k}
\end{align}
At any step of this process, the sites between $k$ and $k+M$ are in the $T$-phase, from the perspective of the domain that is being moved the process could be interpreted as sequential change of sign of $t_j$ from right to left site by site as shown oin Fig. \ref{double_braid_moving}(a). The final form of the Hamiltonian after the entire domain is moved through the region of $+t_j$ is
\begin{align}
  H_{(\text{step }M)} =& \sum_{j=1}^{M} \mu_j H^{(N)}_j + \sum_{j=M+1}^{2M-1} t_j H^{(T)}_j. \label{moving_M}
\end{align}

This completes the adiabatic version of applying a $ U_Z $-gate, where double braids are performed on each site. One might attempt to implement the $U_Z$-gate operation using nanowire of length $M$ instead of $2M$ and just reverse the sign of $t_j$ sequentially site by site. This would be an equivalent approach with the same effect using a shorter nanowire but a disadvantage is the requirement that the sign of $t_j$ would need to be controlled on every site of the chain instead of just single site.

\section{Arbitrary $Z$-rotation and extension to multiple chains}
\label{section_arbitrary_z_rotation_multi}

In the previous section we introduced quantum gates that operate on topological braids implementing the logical $Z$-operation. We described the protocol in both unitary and adiabatic framework by modifying the
coefficients of the Kitaev model Hamiltonian (\ref{hamiltonian_majorana_full}). In this section we extend this theory to implement a $Z$-rotation by an arbitrary angle $\phi$ and extend it to act on multiple topological qubits.


\subsection{The coupler spin}

\begin{figure*}[t]
\includegraphics[width=\linewidth]{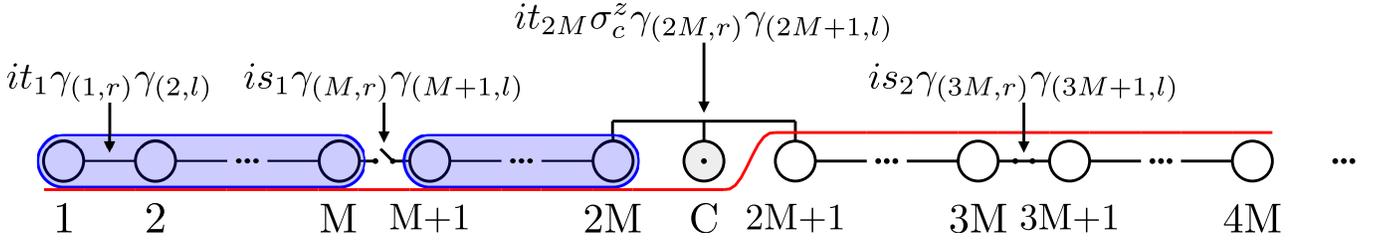}
\caption{A diagram showing Hamiltonian terms associated to four nanowire system, where each nanowire is of length $M$. There are two topological domains present in the system and stored on the two left-most nanowires.}\label{hamiltonian_diagram}
\end{figure*}

We demonstrated earlier that moving the domain through regions of same sign of $t_j$ does not affect the topological state yet moving it through a region with reversed sign of $t_j$ is equivalent to performing a $U_Z$ operation. In order to control the sign of the special region, in Ref. \cite{Backens2017} it was proposed to introduce an extra degree of freedom called the coupler. Assuming the controlled region is coupling sites $M$ and $M+1$ (in a system bigger than $M$) the coupling term takes the following form
\begin{align}
  H_c &= -i t_M \sigma^z_c \gamma_{(M,r)}\gamma_{(M+1,l)}. \label{hamiltonian_coupler}
\end{align}
The term mentioned above corresponds to the central region in Fig. \ref{hamiltonian_diagram} and the region labeled as the ``coupler'' on Fig. \ref{processor_design}.  Potential candidates for the physical implementation of the coupler qubit are either semiconductor quantum dots \cite{press2008complete,bonadeo1998coherent,ishida2013photoluminescence} or superconducting qubits \cite{Backens2017,clarke2008superconducting,barends2014superconducting}.  

Given a topological state $\ket{\psi}$ we can control the $U_Z$ operation in the following way using the coupler degree of freedom.  Denoting the state of the coupler by $\ket{0}_c$ or $\ket{1}_c$, moving the state $\ket{0}_c \otimes \ket{\psi}$ does not affect the state because the sign of coupling term in (\ref{hamiltonian_coupler}) is unchanged. On the other hand, moving the state $\ket{1}_c \otimes \ket{\psi}$ through the coupler region leads to $\ket{1}_c \otimes (-U_Z)\ket{\psi}$ since the sign of the coupling region (\ref{hamiltonian_coupler}) is reversed due to $\sigma^z_c \ket{1}_c = -\ket{1}_c$. Since the $U_Z$ is applied only when the coupler is in the $\ket{1}_c$ state, this effectively implements a controlled-$Z$ operation, where the control qubit is the coupler, and the target qubit is encoded by the MZMs of the chain. The $U_Z$ controlled by the coupler will be denoted as
\begin{align}
  U_Z^c &= \dyad{0}{0}_c + U_Z \dyad{1}{1}_c. \label{controlled_z_definition}
\end{align}

\subsection{Arbitrary $Z$ rotation using the coupler}
\label{section_arbitrary_z_rot}

\begin{figure}[t]
\begin{tabular}{c}
  \includegraphics[width=\linewidth]{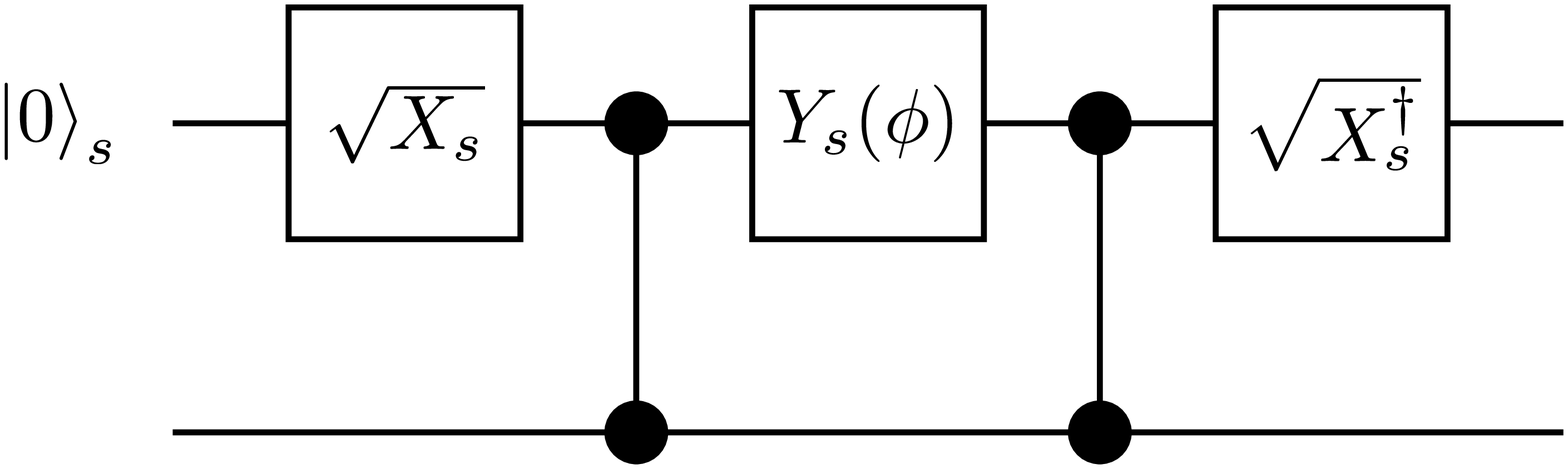} \\
  (a) \\[6pt]
  \includegraphics[width=\linewidth]{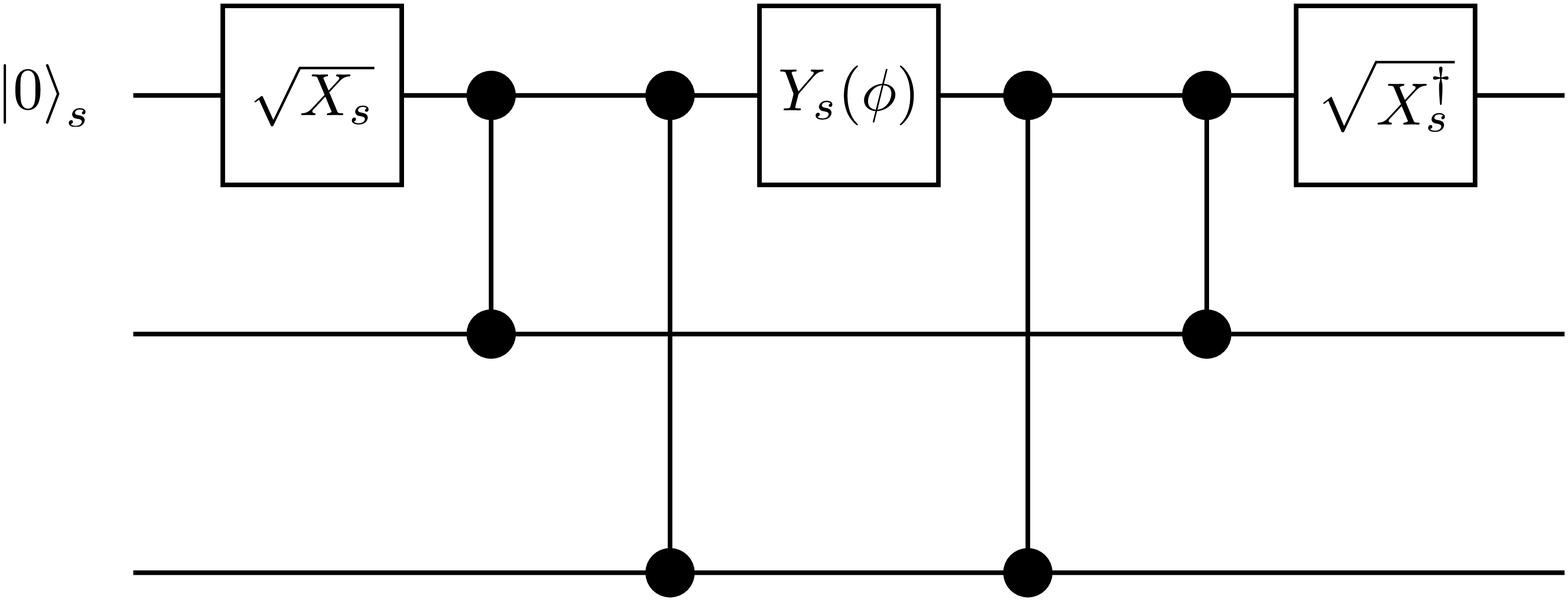} \\
  (b) \\[6pt]
\end{tabular}
\caption{Quantum circuit implementing a $Z_L$-rotation by an arbitrary angle $\phi$ by acting on a coupler which controls the nanowire state $\ket{\psi}$ using through controlled-$Z_L$ operation (a). Extension to multiple logical qubits constructed using same principles by adding additional controlled-$Z_L$ operations (b).}\label{arbitrary_z_circuit}
\end{figure}

So far we have introduced a controlled $U_Z$-gate that is able to process topological qubits built out of topological domain nanowires dependent on the state of the coupler. In addition to being able to control the application of the $U_Z$ operation (\ref{controlled_z_definition}), this in fact makes it possible to implement an arbitrary $Z$-rotation gate according to the following expression
\begin{align}
  U_Z(\phi) &= e^{i \sigma^x_c \pi/2} U_Z^c e^{-i \sigma^y_c \phi/2} U_Z^c e^{-i \sigma^x_c \pi/2} \label{controlled_z}
\end{align}
which can be written equally as the quantum circuit in Fig. \ref{arbitrary_z_circuit}. Consider an arbitrary topological state
\begin{align}
  \ket{\psi^{(0)}}_L &= \alpha \ket{0_L} + \beta \ket{1_L} \label{arbitrary_topological_state}
\end{align}
where $\lvert\alpha\rvert^2+\lvert\beta\rvert^2=1$. Applying the first two coupler gates and the $ U_Z $ we obtain
\begin{align}
  &e^{-i \sigma^y_c \phi/2} U_Z^c e^{-i \sigma^x_c \pi/2}\ket{0}_c\ket{\psi^{(0)}_L}\ket{1_L} \nonumber \\
  =& (\frac{1}{\sqrt{2}}(\cos{\frac{\phi}{2}} \ket{0}_c + \sin{\frac{\phi}{2}} \ket{1}_c) (\alpha \ket{0_L} + \beta \ket{1_L}) \nonumber \\
  +& \frac{i}{\sqrt{2}} (\cos{\frac{\phi}{2}} \ket{1}_c - \sin{\frac{\phi}{2}} \ket{0}_c) (\alpha \ket{0_L} - \beta \ket{1_L})) \label{unitary_z_braiding_states_5}
\end{align}
Performing another controlled $U_Z$ will disentangle the coupler from the domain and the final unitary of the coupler returns it to the initial state.  We thus have
\begin{align}
  & e^{i \sigma^x_c \pi/2} U_Z^c e^{-i \sigma^y_c \phi/2} U_Z^c e^{-i \sigma^x_c \pi/2} \ket{\psi^{(0)}_L} \nonumber \\
  =& \ket{0}_c ( \alpha e^{-i \phi/2} \ket{0_L} + \beta e^{i \phi/2}\ket{1_L} ). \label{unitary_z_braiding_states_8}
\end{align}
We thus see that the logical state has been rotated by an angle $ \phi $ around the $ Z $ axis.   We note that this step only has a partial topological protection.  The part of the circuit involving $ U_Z $ is topologically protected since it involves a non-local operation throughout the topological chain.  However, as can be seen from (\ref{unitary_z_braiding_states_5}), the coupler qubit involves a rotation about an angle $ \phi $, which is eventually applied to the qubit.  Hence if an error occurs during the operation on the coupler qubit, it is susceptible to errors.  We may expect that this operation does not have full topological protection since a rotation about an arbitrary angle corresponds to a non-Clifford gate.  Since only Clifford gates are implementable using braids in this model, the lack of topological protection is the price to be paid for extending the gates beyond the Clifford set. These steps provide foundation for constructing multi-qubit entangling gate $Z^{12}_L(\phi)$.

\subsection{Adiabatic scheme for arbitrary Z rotation}

\begin{figure*}[t]
\begin{tabular}{cc}
  \raisebox{.375cm}{\includegraphics[height=8.5cm]{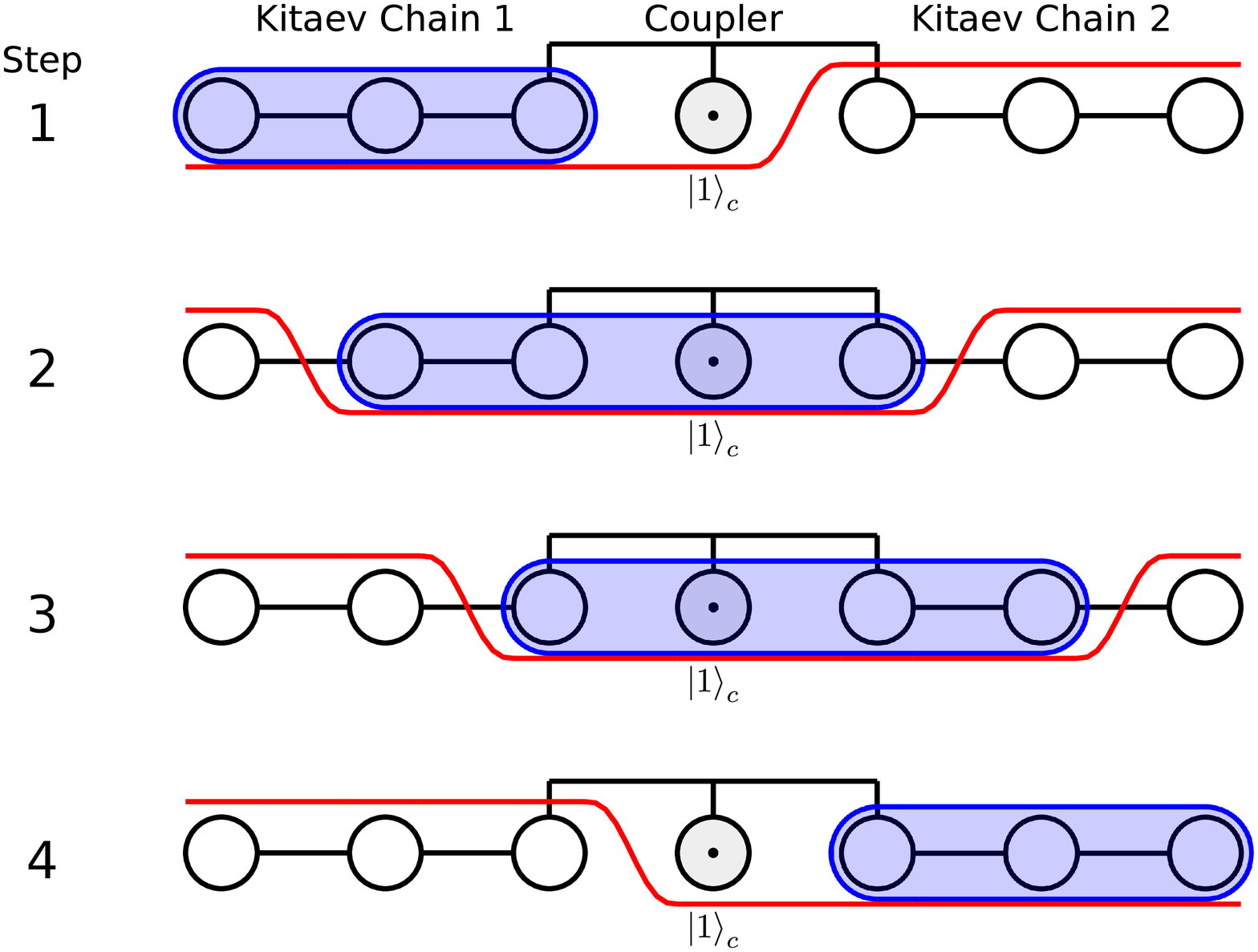}} &
  \includegraphics[height=9cm]{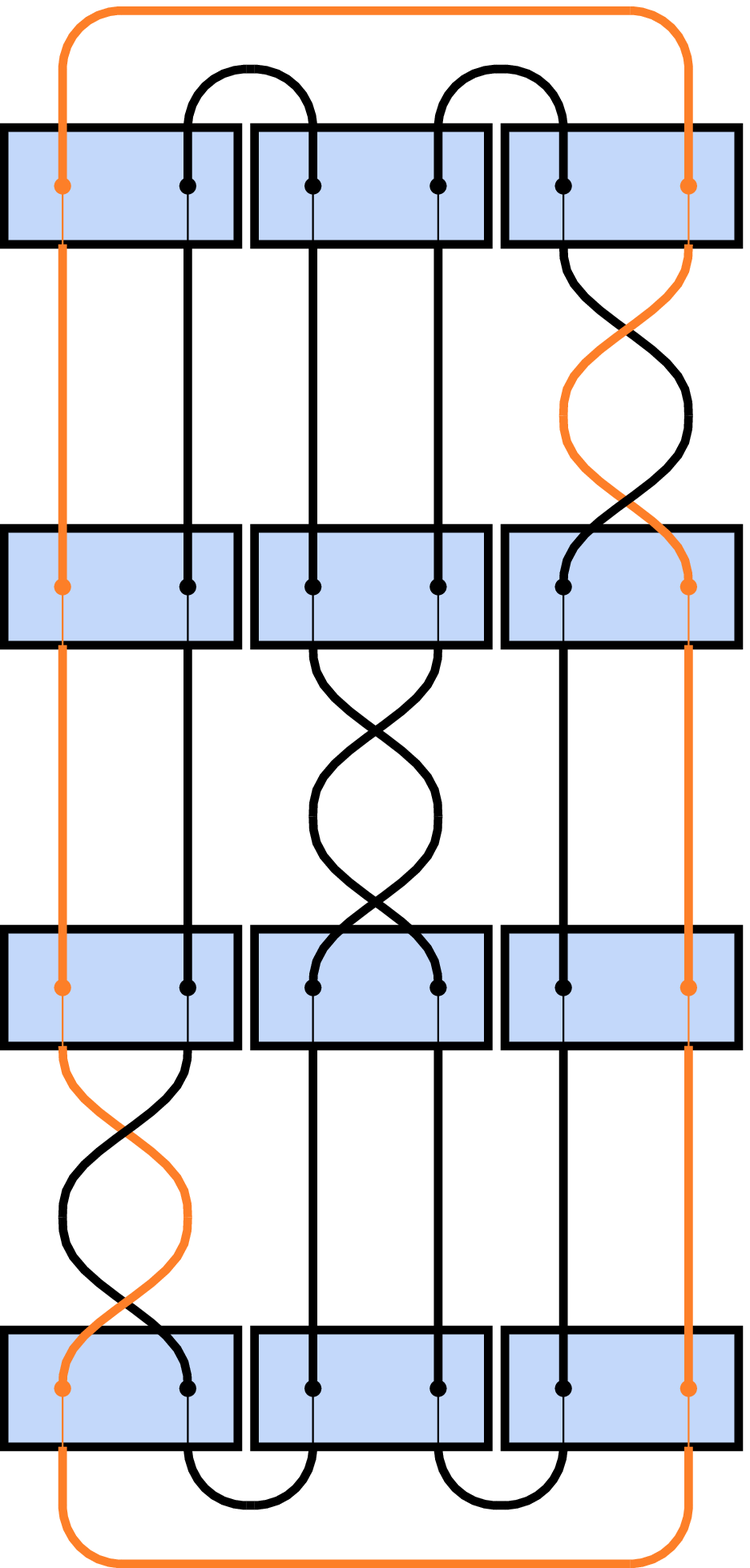} \\
(a) & (b) \\[6pt]
\end{tabular}
\caption{Moving the anyon through the $3$-spin coupling with different states of the coupler (a) and corresponding local braids performed analogously (b) and the quantum circuit controlled by the coupler spin (c)}\label{1d_protocol_vs_braiding_z}
\end{figure*}

The purpose of this section is to demonstrate how to perform the operation described in Sec. \ref{section_arbitrary_z_rot} by adiabatically changing the coefficients of the Hamiltonian (\ref{hamiltonian_fermionic}) thereby implementing an adiabatic protocol equivalent to (\ref{controlled_z}). At each step of the protocol the Hamiltonian is partially in the $N$-phase (\ref{hamiltonian_majorana_trivial}) and partially in the $T$-phase (\ref{hamiltonian_majorana_topological_plus}). The overall approach is the same as that described in Sec. \ref{section_performing_braid_while_moving} except for incorporating the coupler unitaries that appear in (\ref{controlled_z}). Initially, the Hamiltonian takes the form
\begin{align}
  H_{(\text{step }0)} =& \sum_{j=1}^{M-1} t_j H^{(T)}_j  + \sum_{j=M}^{2M} \mu_j H^{(N)}_j.  \label{coupler_moving_0}
\end{align}
Consecutive steps of the protocol sequence can be derived by following the steps from (\ref{unitary_z_braiding_states_5})-(\ref{unitary_z_braiding_states_8}). The protocol will consist of $2M$ adiabatic steps and $3$ unitary steps so in total there would be $2M+3$ states in total and also $2M$ Hamiltonians of the form $H_{(\text{step }k)}$
\begin{align}
  H_{(\text{step }k)} =& \sum_{j=1}^{k} \mu_j H^{(N)}_j + \sum_{\substack{j=k+1\\ j \neq M}}^{M+k} t_j H^{(T)}_j \nonumber \\
  - & t_M \sigma^z_c H^{(T)}_M + \sum_{j=M+k}^{2M} \mu_j H^{(N)}_j. \label{coupler_moving_k}
\end{align}
The system consists of two nanowires coupled with region controlled by the coupler spin based on (\ref{hamiltonian_coupler}). At each step of the protocol, the state is a ground state of the Hamiltonian and matches the corresponding state from the unitary protocol described by (\ref{unitary_z_braiding_states_5})-(\ref{unitary_z_braiding_states_8}). As the controlled-$U_Z$ is applied twice in (\ref{controlled_z}) the domain moves near the coupler twice. First time during the initial sequence when we adiabatically sweep $H_{(\text{step }k)}$ to $H_{(\text{step }k + 1)}$ until $k=M$ is reached. Second time during the returning sequence when we adiabatically sweep $H_{(\text{step }k)}$ into $H_{(\text{step }k-1)}$ until $k=1$ is reached. The three unitary steps are the operations applied to the coupler spin --- putting it in the superposition state, rotating it about the $Y$-axis by an angle $\phi$ and bringing it back to its initial $\ket{0}_c$ state. Adiabatic transitions introduce global phase errors so the states resulting from those adiabatic transitions are equal to (\ref{unitary_z_braiding_states_5})-(\ref{unitary_z_braiding_states_8}) up to a global phase factor.

\subsection{Extension to multiple chains}

The form (\ref{controlled_z}) can be generalized to an entangling-$Z_1 Z_2$ gate involving two nanowires $U_Z^{12}(\phi)$. The entangling gate constructed in this way can even be scaled up to an arbitrary number of qubits. The unitary (\ref{controlled_z}) generalized from a single $T$-phase region up to two $T$-phase regions takes the form
\begin{align}
  U_Z^{12}(\phi) &= e^{i \sigma^x_c \pi/2} U_{Z_1}^c U_{Z_2}^c e^{-i \sigma^y_c \phi/2} U_{Z_2}^c U_{Z_1}^c e^{-i \sigma^x_c \pi/2}. \label{controlled_multi_z}
\end{align}
The equivalent quantum circuit for the two qubit case is shown in Fig. \ref{hamiltonian_diagram}(b). It is straightforward to extend (\ref{controlled_z}) and (\ref{controlled_multi_z}) to involve more topological qubits since the only role of the coupler qubit is to transfer the phase of the rotation to the logical qubits. The overall process follows the steps equivalent to (\ref{unitary_z_braiding_states_5})-(\ref{unitary_z_braiding_states_8}) but this time more logical qubits get entangled with the coupler. Let us consider an arbitrary state involving two logical qubits
\begin{align}
  \ket{\psi^{(0)}_L} &= (\alpha \ket{0_L}_1 + \beta \ket{1_L}_1)(\alpha^\prime \ket{0_L}_2 + \beta^\prime \ket{1_L}_2). \label{general_zz_initial_state}
\end{align}
Preparing the coupler and applying both controlled operations from (\ref{controlled_multi_z}) on state (\ref{general_zz_initial_state}) yields
\begin{align}
  & U_{Z_2}^c U_{Z_1}^c e^{-i \sigma^x_c \pi/2} \ket{0}_c \ket{\psi^{(0)}_L} \nonumber \\
  =& \ket{+i}_c (\alpha \alpha^\prime \ket{00_L}_{12} + \beta \beta^\prime \ket{11_L}_{12}) \nonumber \\
  + & \ket{-i}_c (\alpha \beta^\prime \ket{01_L}_{12} + \beta \alpha^\prime \ket{10_L}_{12})
\end{align}
Then we apply the coupler rotation about the $Y$-axis which creates phase factors under terms entangled with the coupler. Finally, applying the remaining operators of (\ref{controlled_multi_z}) disentangle the logical qubits from the coupler and brings the coupler back to its initial state
\begin{align}
  &U_Z^{12} (\phi) \ket{0}_c \ket{\psi^{(0)}_L} \nonumber \\
  = \ket{0}_c (e^{-i\phi/2}&(\alpha \alpha^\prime \ket{00_L}_{12} + \beta \beta^\prime \ket{11_L}_{12})) \nonumber \\
  + e^{i\phi/2}&(\alpha \beta^\prime \ket{01_L}_{12} + \beta \alpha^\prime \ket{10_L}_{12}))
\end{align}
which is the form we would expect to get after applying a logical-$Z_1 Z_2$ effective Hamiltonian on two logical qubits.  This is an entangling gate and can create entanglement between the topological chains.

\subsection{Adiabatic scheme for $ZZ$ rotation}

We now describe an adiabatic version of the operation introduced in the previous section. Implementing (\ref{controlled_multi_z}) is equivalent to how we implemented (\ref{controlled_z}) but we must ensure the nanowires storing the topological domains are never in contact so that the logical qubits remain distinct at all times.  We introduce modified versions of (\ref{coupler_moving_0})-(\ref{coupler_moving_k}) which connect four nanowires of lengths $M$. Unlike the unitary description for which the order of applying the controlled-$Z_L$ is arbitrary due to the fact they commute, in case of the adiabatic protocol even for commuting terms the order of operations needs to be considered. As the $T$-phase regions are inside of the 1D geometry we cannot entangle them with the coupler in any order, they must pass through the regions involving the coupler one by one and return in reverse order. This is much like a FILO queue (first in, last out). The entire process is summarized in four steps, each being a sequence of $M$ adiabatic transitions characterized by the index $1 \leq k \leq 2M$. 

\begin{figure}[t]
\includegraphics[width=\linewidth]{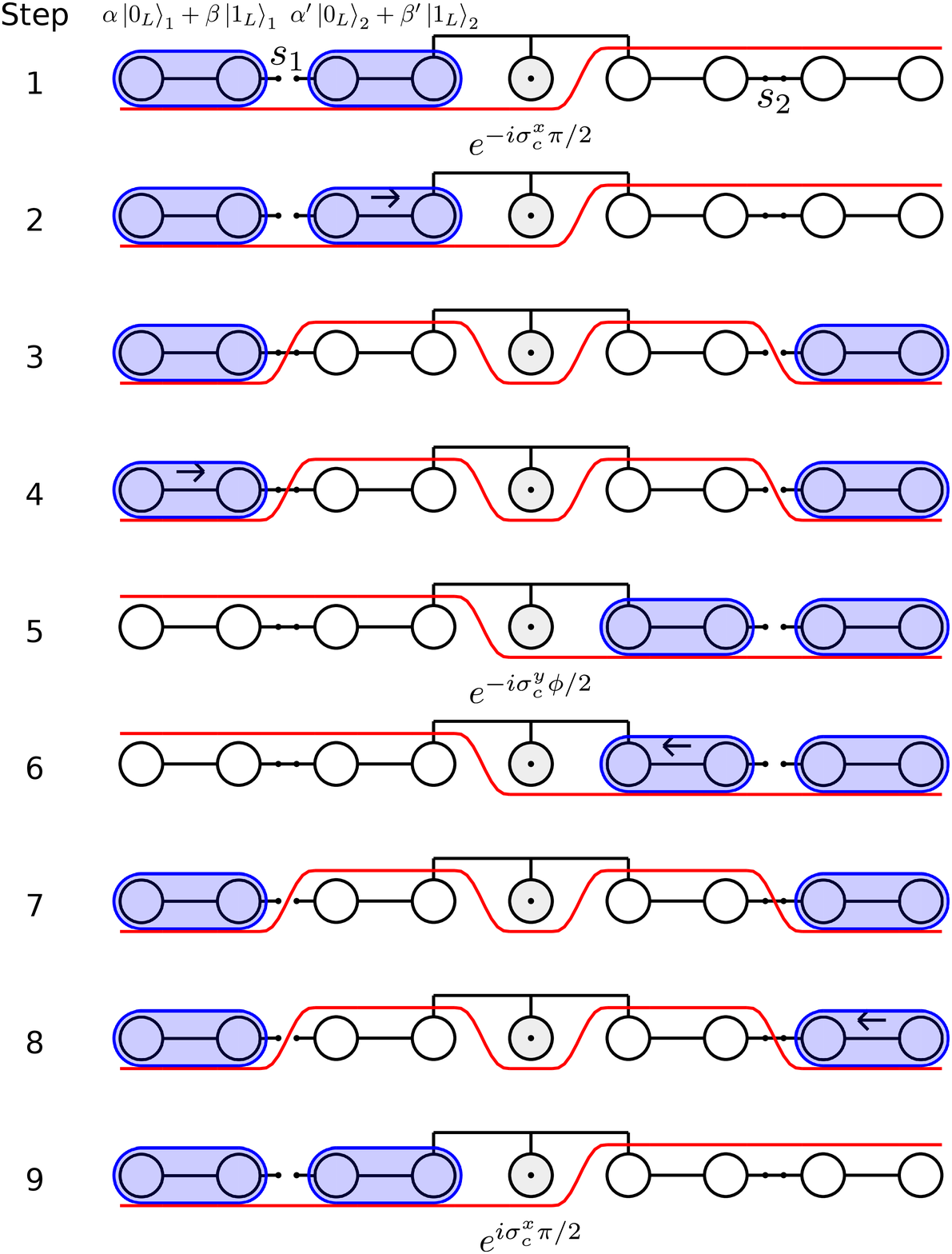}
\caption{Adiabatic scheme to entangle two topological qubits. Each chain is of length $M=2$ and at the end of the sequence a $Z_1 Z_2$ Hamiltonian is implemented. The steps $1$, $5$ and $6$ represent the unitary operations applied to the coupler. Steps $2-4$ and $6-8$ represent moving the topological regions between the chains. Open circles denote fermion sites, filled circle is the coupler, the switch denotes a region with suppressed $t_j$ and $\Delta_j$ such that the tunneling between the sites is decreased, the filled ovals denote the topological regions, and the curved line gives the value of $\mu_j$ for the site.}\label{1d_protocols}
\end{figure}

We provide a diagram showing the the sequence of adiabatic steps for case of $N=2$ and $M=2$ in Fig.  \ref{1d_protocols}. Step $1$ shows us the initial configuration of the system and performs the unitary operation preparing the coupler in the superposition state. As shown on the diagram at step $1$, both $T$-phase domains are located on the left relative to the coupler. Step $2$ consists of moving the right topological domain to the other side of the coupler. In step $3$, the couplings are changed to ensure separation between the domains before we move leftmost domain, which is done in step $4$. Step $5$ consists of unitary coupler rotation. Remaining steps $6$-$8$ are the same operations as steps $2$-$4$ but applied in reverse order. The final step $9$ is a unitary operation that brings the coupler back to its initial state, reversing the operation performed in step $1$. This completes the entire protocol for two logical qubits.

We provide two Hamiltonians, one for each topological region that is moved through the coupler.  The first step is to move the topological region on the immediate left of the coupler to the right, as can be seen in Fig. \ref{1d_protocols}.  This is achieved by the Hamiltonian
\begin{align}
  H_{(\text{step }k)}^{(1)} = &\sum_{j=1}^{M-1} t_j H^{(T)}_j + \sum_{j=M}^{M+k} \mu_j H^{(N)}_j \nonumber \\
  -t_M \sigma^z_c H^{(T)}_{2M} + &\sum_{\substack{j=M+k\\j\neq 2M}}^{2M+k} t_j H^{(T)}_j + \sum_{j=2M+k}^{4M} \mu_j H^{(N)}_j \label{zz_hamiltonian_1}
\end{align}
which moves the rightmost domain through the coupler. The coupler is coupled to sites $2M$, $2M+1$ as two domains in the system are present. The second Hamiltonian moves the topological domain on the far left to the right side of the coupler (Fig. \ref{1d_protocols}) and is written
\begin{align}
  H_{k}^{(2)} = &\sum_{j=1}^{k} \mu_j H^{(N)}_j  + \sum_{\substack{j=k\\j\neq 2M}}^{3M-2} t_j H^{(T)}_j\nonumber \\
  - t_M \sigma^z_c H^{(T)}_{2M} + &\sum_{j=2M+k}^{3M} \mu_j H^{(N)}_j + \sum_{j=3M}^{4M-1} t_j H^{(T)}_j \label{zz_hamiltonian_2}
\end{align}
The way Hamiltonians (\ref{zz_hamiltonian_1})-(\ref{zz_hamiltonian_2}) are constructed prevents the domains from interacting as the topological domains are always stored on separate uncoupled nanowires. Whether neighboring nanowires are coupled or not is represented by value $s_j \in \{0, 1\}$, for a four nanowire system coupling is controlled by two such parameters $s_1$ for the nanowires on the left side of the coupler and $s_2$ for the nanowires on the right side of the coupler as labeled on the top row in Fig. \ref{1d_protocols}.

\section{Arbitrary qubit rotation}
\label{section_arbitrary_qubit_rotation}

Until now we have developed the theoretical tools for performing effective $Z$ and $Z_1 Z_2 $ Hamiltonians. In order to perform universal quantum computing, we additionally require single logical qubit rotations around another axis in addition to $Z$. In this section we discuss how the conversion operator $U_c$ (\ref{conversion_operator}) can be used to implement an rotation about an arbitrary axis on a topological qubit. The general idea of the approach will be to unbraid the state of a single chain such that the logical information is on a single site, on Fig. \ref{processor_design} marked as the ``local gate''. The arbitrary rotation is applied on this site, then the information is put back in the MZM modes by re-braiding the information. This breaks the topological protection of the state. However, this allows us to perform an arbitrary single qubit logical gate which may be useful to perform more general quantum operations in combination with topologically protected gates.

\subsection{Unitary formulation}

Let $U^{T}_{\bm{u}}(\phi)$ be a general qubit rotation of the topological qubit acting on logical space state of the form $\alpha \ket{0_L} + \beta \ket{1_L}$ where $\ket{0_L}$ and $\ket{1_L}$ are states describing logical space of $T$-phase of length $M$. Now let $U_{\bm{u}}(\phi)$ be an equivalent operation to $U^{T}_{\bm{u}}(\phi)$ but acting on an equivalent state of a single site in the $N$-phase which takes the general form form $\alpha \ket{0}_N + \beta \ket{1}_N$. A general form of such rotation can be constructed using the three rotations about each of the Bloch sphere axes. Our goal here is to relate $U^{T}_{\bm{u}}(\phi)$ and $U_{\bm{u}}(\phi)$ operations. This is achieved using the conversion operator $U_c$ (\ref{conversion_operator}) in the following way
\begin{align}
  U^{T}_{\bm{u}}(\phi) &= U_c U_{\bm{u}}(\phi) U_c^\dagger. \label{non_topo_single_qubit_rotation}
\end{align}
An arbitrary rotation can be defined using above operators and parametrized by a unit vector $\bm{u} = (u_x, u_y, u_z)$. This helps us define a general form of $U_{\bm{u}}(\phi)$ to be
\begin{align}
  U_{\bm{u}}(\phi) &= \exp[-i\frac{\phi}{2}(u_x X_n + u_y Y_n + u_z Z_M)] \label{single_site_gate}
\end{align}
where $X_n$, $Y_n$ and $Z_M$ are Pauli operators acting on site $n$. We can demonstrate how $U^{T}_{\bm{u}}(\phi)$ works by expanding it and applying property (\ref{uc_effect}) of the $U_c$ conversion operator
\begin{align}
  U^{T}_{\bm{u}} (\alpha \ket{0_L} + \beta \ket{1_L}) =& U_c U_{\bm{u}}(\phi) (\alpha \ket{0}_N + \beta \ket{1}_N) \\
  =& U_c (\alpha^\prime \ket{0}_N + \beta^\prime \ket{1}_N) \\
  =& \alpha^\prime \ket{0_L} + \beta^\prime \ket{1_L}
\end{align}
where $\alpha^\prime, \beta^\prime$ are the rotated coefficients of $\alpha, \beta$. This shows that $U^{T}_{\bm{u}}(\phi)$ implements a rotation about an arbitrary axis parametrized by unit vector $\bm{u}$ and acts on the topological space.

The advantage of using the phase transition is it allows us to perform any logical operation on the topological domain. It could also be potentially scaled up to two topological domains if we localize the quasifermions into fermions that are physically close to each other allowing them to create an interaction between them. The disadvantage is lack of topological protection as the domain is almost entirely destroyed during the operation.

\subsection{Adiabatic scheme}

The adiabatic protocol associated to $U^{T}_{\bm{u}}(\phi)$ (\ref{non_topo_single_qubit_rotation}) follows directly from its unitary counterpart. As $U^{T}_{\bm{u}}(\phi)$ is defined in terms of the $U_c$ conversion operator and the single site gate operation $U_{\bm{u}}(\phi)$ the protocol requires us to define those operations in a way that applies to the nanowire system Hamiltonian (\ref{hamiltonian_majorana_full}). The adiabatic protocol for the $U_c$ operation has been described in Sec. \ref{section_conversion_ground_states_braiding}. We do not describe physically how the single qubit rotation (\ref{single_site_gate}) would be performed, since this depends upon the choice of the physical qubit, and these are well developed. An arbitrary single qubit operation can be executed in a suitable way depending on the implementation of the qubit at the region labeled as ``local gate'' on Figure \ref{processor_design}. 

\begin{figure}[t]
\centering
\begin{tabular}{cc}
    \raisebox{.375cm}{\includegraphics[height=12cm]{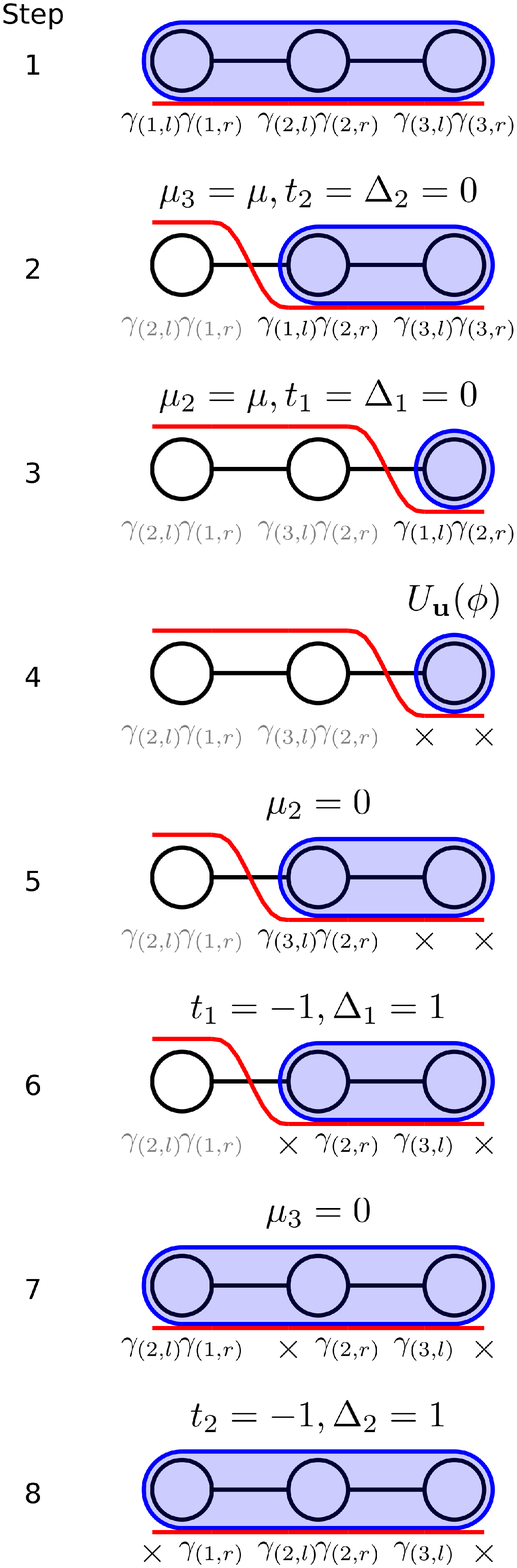}}
  &
  \includegraphics[height=12.75cm]{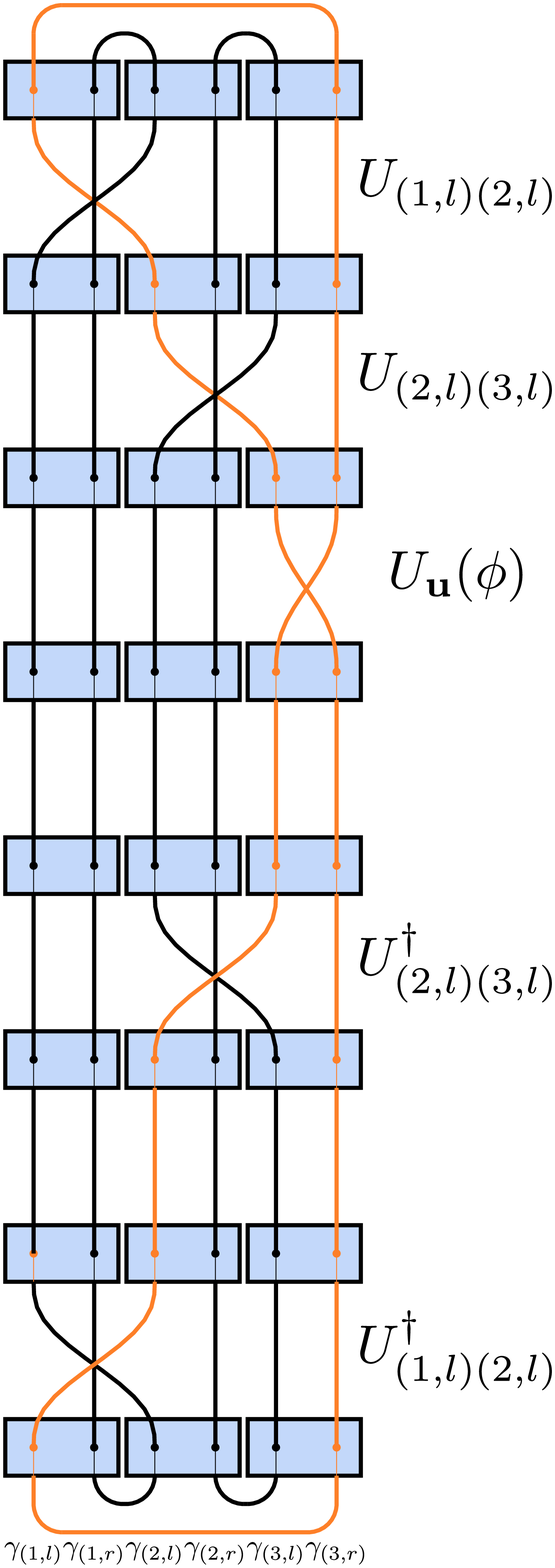}
  \\
(a) & (b) \\[6pt]
\end{tabular}
\caption{The adiabatic protocol (a), sequence of local braids or a unitary protocol (b) both implementing $U^{T}_{\bm{u}}(\phi)$ operation.}\label{1d_protocol_vs_braiding_x}
\end{figure}

We also show how the adiabatic protocol and its braiding counterpart compare in Fig. \ref{1d_protocol_vs_braiding_x} (a) for the case of $M=3$. Step $1$ shows the initial configuration of the nanowire, which is entirely in the $T$-phase. Steps $2$ to $3$ destroy the $T$-phase by creating $N$-phase from the left side. This shrinks the topological qubit changing the quasifermion into a regular fermion localized on the rightmost site. Step $4$ applies the required rotation to the rightmost site. Steps $5$ to $9$ turn the $N$-phase sites back into the $T$-phase sites extending the domain until it spreads over entire nanowire again. Fig. \ref{1d_protocol_vs_braiding_x}(b) shows the corresponding steps by visualizing the unitary braids.

\section{Numerical simulation}

We performed a series of numerical tests of the protocols described earlier. Using QuTip Quantum Toolbox library \cite{JOHANSSON20131234, JOHANSSON20121760} we implemented the Hamiltonian (\ref{hamiltonian_majorana_full}) and numerically simulated adiabatic transitions corresponding to protocols associated to quantum gates described in this work.

\subsection{Energy fluctuation study}

The motivation of studying fluctuations in energy is to confirm topological protection of our gate operations. Such a study is relevant to evaluate gate sensitivities to errors as topological states are protected by the bulk energy gap as demonstrated in Ref. \cite{kitaev2001unpaired}. In particular, we numerically compare $U_Z(\pi)$ and $U^{T}_{\bf{u_x}}(\pi)$. Neither of those operations are topologically protected, yet we would expect the energy to fluctuate to a much larger extent in the case of $U^{T}_{\bf{u_x}}(\pi)$ as this operation is destroying the the $T$-phase completely localizing the quasifermion.

\begin{figure}[t]
\includegraphics[width=\linewidth]{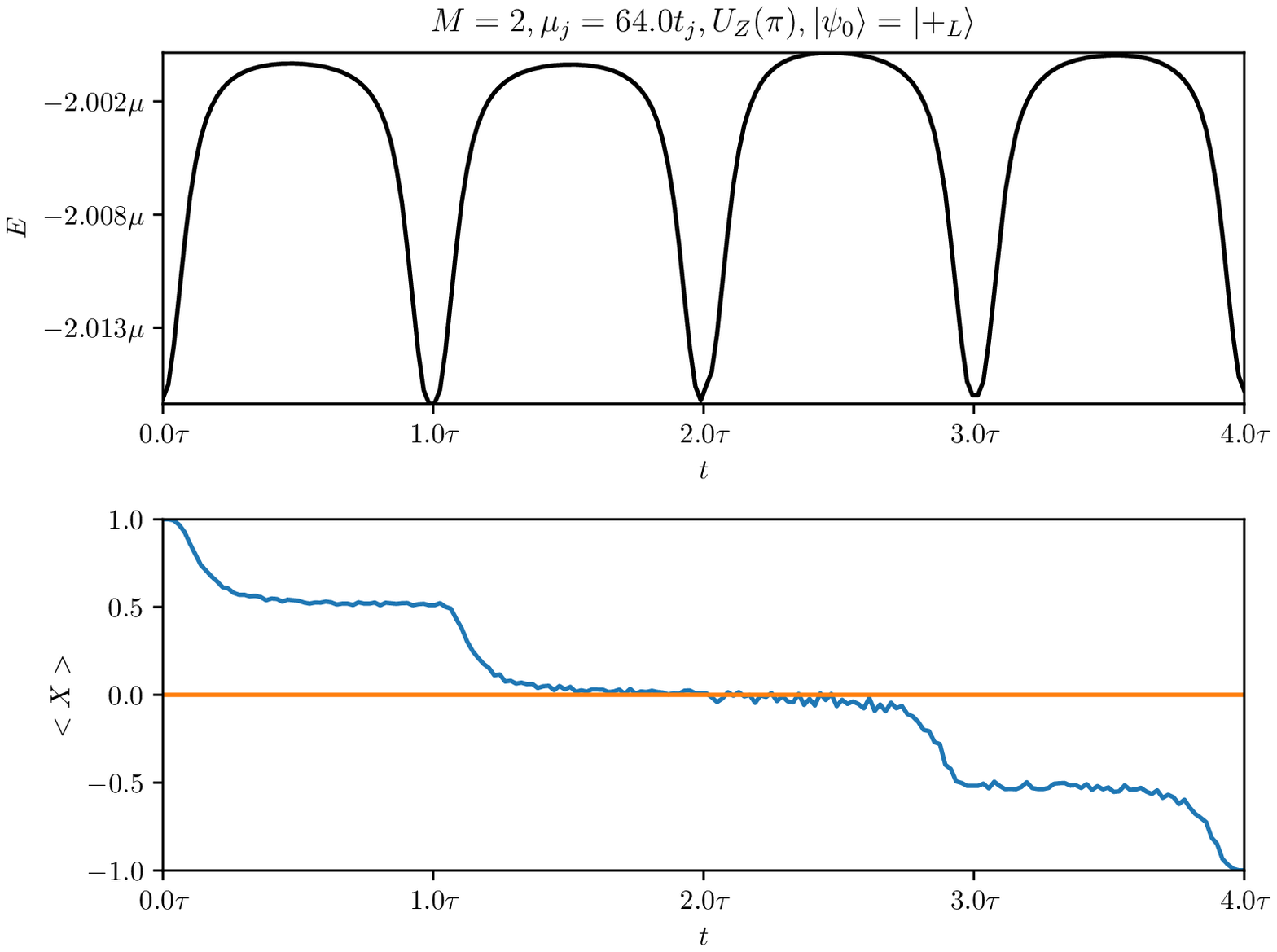} \\
\includegraphics[height=0.75cm]{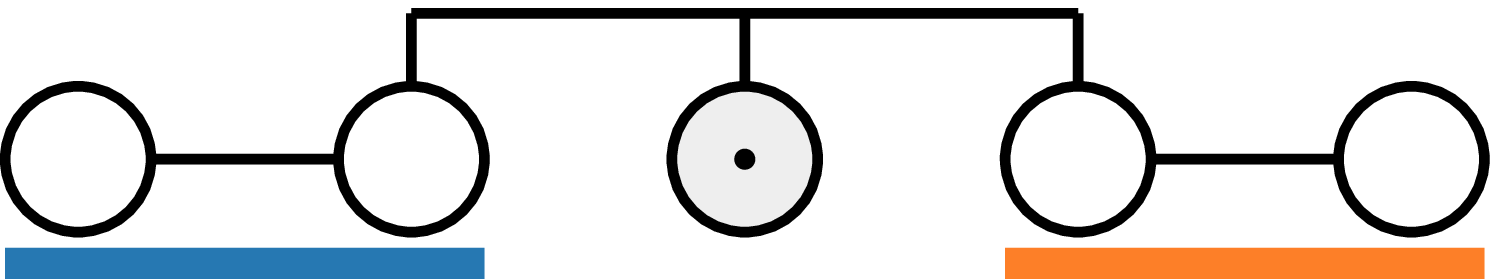} \\
\caption{Energy and the spin expectation value of the state plots for the $U_Z(\pi)$-braiding. On the bottom a diagram is provided picturing color coding legend explaining how to associate colors of $\expval{Z}$ plot to positions in the nanowire.}\label{energy_plots_z}
\end{figure}

\begin{figure}[t]
\includegraphics[width=\linewidth]{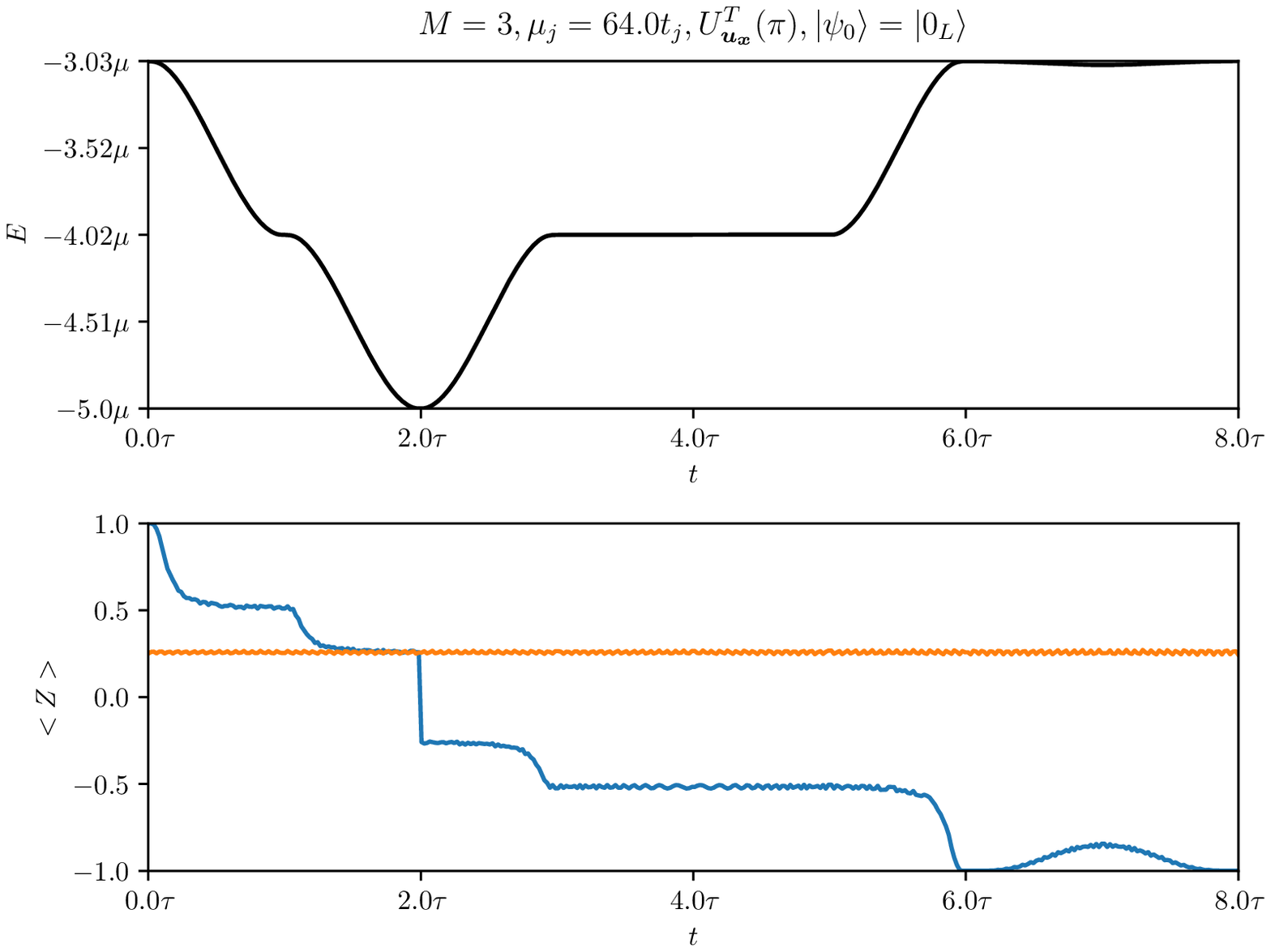} \\
\includegraphics[height=0.75cm]{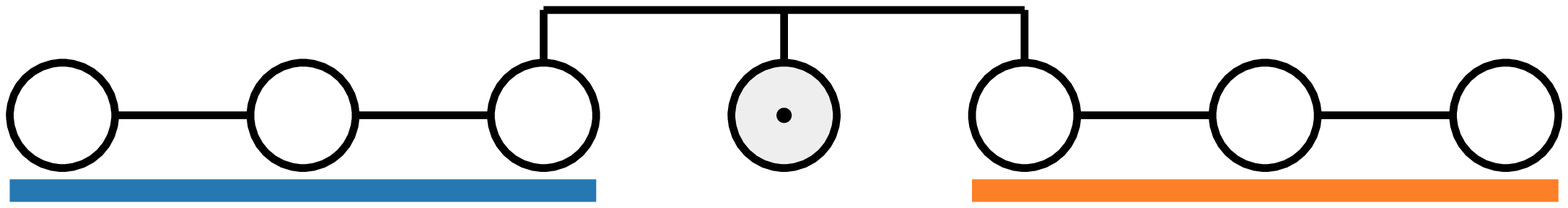} \\
\caption{Energy and the spin expectation value of the state plots for the $U^{T}_{\bf{u_z}}(\pi)$-braiding. On the bottom a diagram is provided picturing color coding legend explaining how to associate colors of $\expval{X}$ plot to positions in the nanowire.}\label{energy_plots_x}
\end{figure}

We performed a numerical study simulating the adiabatic protocols associated with the $U_Z(\phi)$-gate initialized in the $\ket{+_L}$ state as shown in Fig. \ref{energy_plots_z}, and the $U^{T}_{\bf{u_x}}(\pi)$-gate initialized to the $\ket{0_L}$ state as shown in Fig. \ref{energy_plots_x}. Each of the figures contains two plots, the upper plot shows overall energy of the system measured by numerically computing the expectation value of the Hamiltonian $H(t)$ at time step $t$ of the state $\ket{\psi_t}$. The lower plot demonstrates the correctness of the performed operation, by numerically taking the expectation value of the relevant operator such that the initial state of the system is a $+1$ eigenvalue eigenstate of that operator. For the energy study of the $U_Z(\phi)$-gate with $\ket{+_L}$ as the initial state, we measure $X =\dyad{+_L}{+_L} - \dyad{-_L}{-_L}$ and for the $U^{T}_{\bf{u_x}}(\pi)$-gate with $\ket{0_L}$ as the initial state we measure $Z = \dyad{0_L}{0_L} - \dyad{1_L}{1_L}$. While performing the numerical simulations Fig. \ref{energy_plots_z} and Fig. \ref{energy_plots_x} we set the adiabatic sweep $\tau = 16\pi$ and adiabatic errors are the only source of errors considered in the simulation. No noise nor decoherence model has been implemented.

Our results show that the energy of $U^{T}_{\bf{u_x}}(\pi)$-gate fluctuates by the order of $(M-1)\mu$ which is as expected as the topological domain shortens by $M-1$ sites transforming the chain into the $N$-phase with each site of energy $\mu$. This is can be observed on Fig. \ref{energy_plots_x}. The case of the $U_Z(\phi)$-gate shows the energy variations of a much larger magnitude, of order $t$. This is is expected due to the fact that whenever a $T$-phase gets destroyed on one site it also becomes extended on another while moving. Both cases show the gate gets applied correctly by changing the initial state from the $+1$ eigenstate into the $-1$ eigenstate.

\subsection{Fidelity study}

\begin{figure*}[ht]
\includegraphics[width=\textwidth]{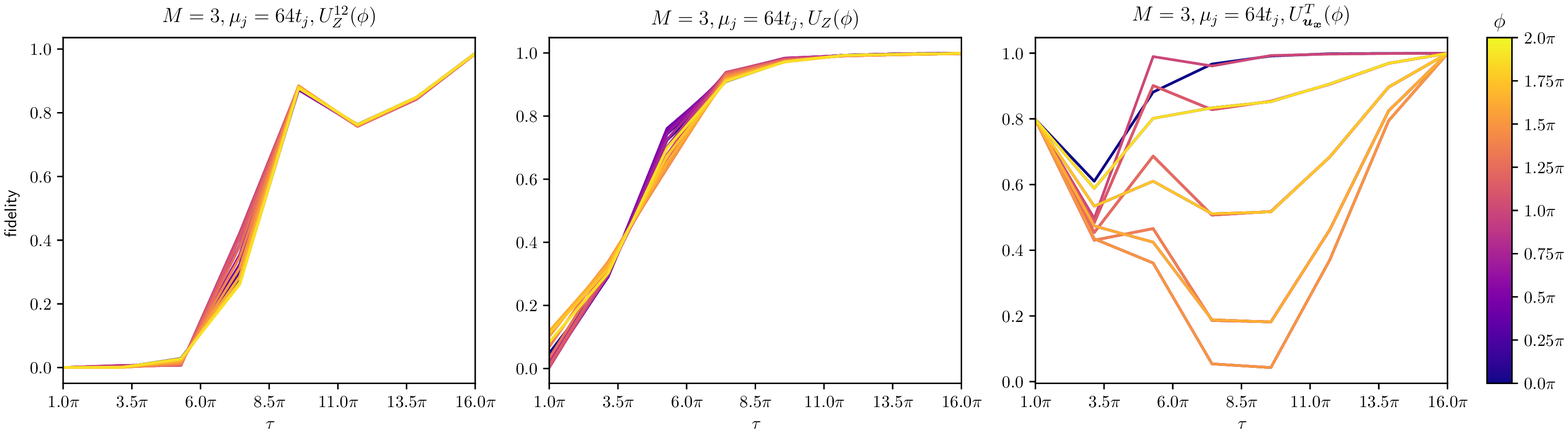}
\caption{Fidelity plots versus adiabatic time constant obtained after numerically simulating entire protocol for all three gates $U_Z^{12}(\phi)$, $U_Z(\phi)$ and $U^{T}_{\bf{u}}(\pi)$ with $\bf{u}=\bf{u_x}$. Fidelity is calculated for a series of angles $\phi$ ranging over $16$ evenly distributed angles $\phi$ from the unit circle.}\label{fidelity_plots}
\end{figure*}

A more direct measure of the success of the adiabatic gates is to study the fidelity of the state as a function of the adiabatic sweep time $\tau$. The results are shown in Fig. \ref{fidelity_plots}. Performing adiabatic transitions too rapidly will lead the system to populate one of the excited states, which is visible on our fidelity plots by the curves not reaching value of $1$ for lower values of the adiabatic sweep $\tau$. The curves on each plot differ in the angle of the qubit rotation. The fidelity study was performed for the gates $U_Z^{12}(\phi)$, $U_Z^{12}(\phi)$ and $U^{T}_{\bf{u_x}}(\phi)$ where the angle $\phi= 2\pi/16$, where $n \in \{0, 15\}$.

We considered $T$-phase regions of length $M=3$ and parametrized the topological phase of our system as $\mu=64 t_H$. For each of the gates $U_Z^{12}(\phi)$, $U_Z^{12}(\phi)$ and $U^{T}_{\bf{u_x}}(\phi)$ the system was initialized in state $\ket{+_L}$, $\ket{++_L}$ and $\ket{0_L}$, respectively. We varied the adiabatic sweep time $\tau$ from $\pi$ up to $16\pi$ observing all gates approaching fidelity of $1$ indicating the tested gates perform their computation correctly once the adiabatic regime is achieved.


\section{Summary and conclusions}

We have proposed and analyzed methods for performing quantum gates in a 1DTS in a purely one dimensional geometry.   We provided schemes for performing a $ Z$ operation ($U_Z$), arbitrary $Z$-rotations ($ U_Z(\phi)$), and two-qubit  $Z$-rotations $ U_Z^{12}(\phi)$ on MZM encoded logical qubits.
The $U_Z$-gate is a fully topological gate, while the $U_Z(\phi)$ and $U_Z^{12}(\phi)$ only provides partial topological protection, since the coupler qubit is not protected against errors.  The more general single qubit logical gate does not offer any topological protection but it is required for universal quantum computation. Each of the  logical quantum gates proposed is described in both the unitary language based on sequences of topological braids as well as adiabatic variations of the Kitaev model Hamiltonian. We also described the phase transition between the topological $T$-phase and normal $N$-phase in terms of sequences of braids within the nanowire and successfully associated the result with expected variation of Kitaev model parameters. The unitary formalism provides a more intuitive framework to understand the nature of the gates and map them to the associated adiabatic Hamiltonians. We have confirmed the correctness of this work by performing a series of numerical simulations which studied the energy fluctuation of the entire nanowire while particular gates are performed. This showed the expected energy variation during the process of phase transition between topologically trivial and topological regime. The adiabatic scheme was verified to have a high fidelity of for parameters sufficiently in the adiabatic regime.  

One of the challenges of this platform is the limited set of operations that are present using topologically protected operations.  A fundamental limitation is that the type of anyon in the Kitaev chain is of the Ising type, which means that only Clifford gates can be produced by braiding.  This means that additional, non-topologically protected operations must necessarily be included to perform universal quantum computing. In this study, we used a coupler-based approach, which was motivated as an alternative to the T-junction approach, but also provides a natural way of introducing non-Clifford gates.  Recently performing high-fidelity single qubit quantum gates has become feasible in numerous systems  \cite{ladd2010quantum,huang2020superconducting}. As a hybrid system, the role of the topologically protected qubits could serve as storage medium taking a similar role to quantum memories, where quantum information can be stably stored for relatively long times.  When the quantum information needs to be manipulated, it can be done so using topologically unprotected, or partially topologically protected methods, as introduced in this paper. Using such a hybrid approach may be an effective way of combining the expected stability of topological quantum state encodings, with the controllability of existing qubit systems. 


\bibliographystyle{apsrev}
\bibliography{references}

\end{document}